\documentclass[10pt,journal,compsoc]{IEEEtran}
\newif\ifpeerreview

\peerreviewfalse

\usepackage[nocompress]{cite}
\usepackage{url}
\usepackage{amsmath,amssymb,graphicx}
\usepackage{xcolor}

\usepackage{times}
\usepackage{epsfig}
\usepackage{graphicx}
\usepackage{amsmath}
\usepackage{amssymb}
\usepackage[]{subfigure}
\usepackage{array,multirow,graphicx}
\usepackage[normalem]{ulem}
\usepackage[bookmarks=false]{hyperref}
\usepackage{array}
\usepackage{makecell}
\usepackage{diagbox}
\usepackage[switch]{lineno}

\usepackage{lipsum} 

\newif\ifdebugon
\debugontrue

\ifdebugon
\newcommand{\delete}[1]{\textcolor{green}{\sout{#1}}}
\else
\newcommand{\delete}[1]{}
\fi

\ifdebugon

\else

\fi

\ifdebugon

\else

\fi

\ifdebugon

\else

\fi

\newcommand{\paperID}{22}

 \title{Learning a Probabilistic Strategy \\ for Computational Imaging  Sensor Selection}
\author{He~Sun,~\IEEEmembership{Member,~IEEE,}
        and~Adrian~V.~Dalca,~\IEEEmembership{Member,~IEEE,}
        and~Katherine~L.~Bouman,~\IEEEmembership{Member,~IEEE}
\IEEEcompsocitemizethanks{\IEEEcompsocthanksitem 
H. Sun and K.L. Bouman are with the Department
of Computing and Mathematical Sciences, California Institute of Technology, CA.\protect\\
E-mail: hesun@caltech.edu
\IEEEcompsocthanksitem A.V. Dalca is with Harvard Medical School and the Massachusetts Institute of Technology, MA.}
}

\begin{document}

\IEEEtitleabstractindextext{%
\begin{abstract}
Optimized sensing is important for computational imaging in low-resource environments, when images must be recovered from severely limited measurements. In this paper, we propose a physics-constrained, fully  differentiable, autoencoder that learns a probabilistic sensor-sampling strategy for optimized sensor design. The proposed method learns a system's preferred sampling distribution that characterizes the correlations between different sensor selections as a binary, fully-connected Ising model. The learned probabilistic model is achieved by using a Gibbs sampling inspired network architecture, and is trained end-to-end with a reconstruction network for efficient co-design. The proposed framework is applicable to sensor selection problems in a variety of computational imaging applications. In this paper, we demonstrate the approach in the context of a very-long-baseline-interferometry (VLBI) array design task, where sensor correlations and atmospheric noise present unique challenges. We demonstrate results broadly consistent with expectation, and draw attention to particular structures preferred in the telescope array geometry that can be leveraged to plan future observations and design array expansions.

\end{abstract}

\begin{IEEEkeywords} 
Computational Imaging, Optimized Sensing, Ising Model, Deep Learning, VLBI, Interferometry
\end{IEEEkeywords}
}

\maketitle
\thispagestyle{empty}

\IEEEraisesectionheading{
  \section{Introduction}\label{sec:introduction}
}
%
%
%
%
\IEEEPARstart{C}{omputational} imaging systems tightly integrate algorithm and sensor design, making it possible to observe phenomena previously difficult or impossible to see with traditional sensors. A common constraint in these imaging systems is that they must operate in low-resource environments. For instance, the measurements collected are severely limited due to radiation dosage in computed tomography (CT)~\cite{kobler2018variational}, speed in Magnetic Resonance Imaging (MRI)~\cite{lustig2008compressed} and microscopy~\cite{rivenson2017deep}, and cost in very-long-baseline-interferometry (VLBI)~\cite{bouman2016computational}. 
Imaging methods are then designed to intelligently fill in the gaps of missing information in order to recover the targeted image.
We propose a framework to jointly \textit{learn} the ideal sampling structure of non-linear, correlated measurements simultaneously with the image recovery procedure, for the purpose of sensor design. We apply this framework to study optimal VLBI telescope array geometries in different observing conditions.


A computational imaging system typically consists of two primary components: the sensing system and the image reconstruction method. The sensing system is most often described by a physics-based forward model $y = f(z)$, which defines how the selected measurements, $y$, are related to the underlying image, $z$. These measurements often experience noise and do not contain enough information on their own to fully specify $z$. Therefore, imaging methods are designed to recover an image, $\hat{z}$, from the observed measurements by imposing pre-defined assumptions, such as image priors or hard physical constraints.

In addition to developing image reconstruction methods, optimization of the sensor is critical for producing high-quality reconstructions.
In many applications the placement (i.e., sampling) of sensors controls the nature of information gathered in measurements. It is especially important that this information be optimized for a specific task when the measurement space can't be fully observed. Previous work in 3D reconstruction~\cite{olague2002optimal}, microscopy~\cite{tian2015computational, kellman2019data}, and astronomical imaging\cite{woody2001radio, boone2001interferometric} have proposed strategies to optimize sensor placement to achieve maximally informative measurements. Seminal work in compressed sensing showed that, under particular sampling distributions, it is possible to recover a sparse signal with near perfect accuracy~\cite{compressedsensing}; this idea revolutionized a number of real-world applications, including MRI~\cite{mricompressedsensing}. However, these theoretical results only hold under certain assumptions, such as a linear forward model or limited noise. 
Consequently, these sampling strategies are surely under-optimized and could be improved in complex imaging systems.

The optimal sensing problem is typically idealized and considered independently from the image reconstruction problem. However, these two problems are intimately related: existing imaging algorithms inform what data we are capable of interpreting and therefore should collect, and the particular data we measure informs what imaging algorithms we should use. Inspired by this insight, we address co-designing the sensor-sampling strategy and the image reconstruction algorithm. The complexity of real-world measurements makes it difficult to tackle this co-design problem analytically. Therefore, we propose to solve a joint optimization problem using deep learning, where the optimal sampling policy is represented as a trainable joint probability distribution over candidate sensors.
The proposed framework enables 1) tackling complicated sampling problems where measurements are correlated and experience substantial noise, 2) exploring multiple plausible sampling strategies simultaneously rather than focus on a single deterministic design, and 3) qualitatively understanding the importance of each sensor and correlation among sensors.

The content of the paper is organized as follows. In Sec.~\ref{sec:background}, we review related work on reconstruction algorithms and optimal sensing. In Sec.~\ref{sec:method}, we present our method that jointly optimizes the sensor-sampling pattern and image reconstruction. In Sec.~\ref{sec:result}, we demonstrate our method on the astronomical imaging VLBI array design problem to identify which telescopes are most important to observe with in different conditions. In Sec~\ref{sec:summary}, we summarize our work and discuss additional sensor-selection applications.

\section{Related Work} \label{sec:background}

\subsection{Image Reconstruction Techniques}

\subsubsection{Regularized Inverse Model}

A common approach taken in computational imaging is to reconstruct the target image from limited sensor measurements by solving a regularized inverse problem:
	\begin{equation} \label{eq:ci}
	\hat{z} = \arg \min_{z} \left[ \mathcal{L}(y , f(z)) + \alpha \mathcal{R} (z) \right],
	\end{equation}
	where $f(z)$ is the forward model of the underlying imaging system, which produces measurements $y$ from the true image $z$, $\mathcal{L}(\cdot)$ is inspired by the data negative log-likelihood, $\mathcal{R}(\cdot)$ is a regularization function, $\alpha$ is the coefficient balancing the two terms, and $\hat{z}$ is the reconstructed image. The $\mathcal{L}(\cdot)$ and $\mathcal{R}({\cdot})$ terms quantify the image's fit with the measurements and the prior knowledge, respectively. The regularization term shrinks the solution domain, consequently making the solution unique even with an incomplete set of measurements. Typically, this inverse problem can be solved using iterative optimization algorithms when explicit image regularizer functions are imposed, such as total variation (TV)\cite{bouman1993generalized}, maximum entropy (MEM)\cite{skilling1984maximum}, sparsity in Fourier or wavelet transformed domain\cite{candes2007sparsity}, or locally-learned distributions\cite{bouman2018reconstructing}.

\subsubsection{Learned Inverse Model}

Deep learning techniques have also been used for computational imaging reconstruction by either directly training a deep reconstruction neural network~\cite{sinha2017lensless} or inserting deep learning ``priors" into model-based optimization frameworks~\cite{venkatakrishnan2013plug}. Unlike many classic computer vision tasks, such as face or object recognition, computational imaging problems are fundamentally related to the physics of a known imaging system. Therefore, recent deep learning based reconstruction methods focus on incorporating known physics to improve computational efficiency, reconstruction accuracy, and robustness. For example, this can be done by adding physics-informed regularization to the neural network loss function~\cite{chen2018reblur2deblur}, only learning the residuals between ground truth and physics-based solutions~\cite{jin2017deep}, or unrolling the physics-based optimization procedure as a neural network~\cite{diamond2017unrolled, hammernik2018learning}.

\subsection{Sensor-Sampling Strategies}

\subsubsection{Analytic Strategies}

Conventionally, sparse sensor sampling is achieved by adopting the Nyquist-Shannon sampling theorem. For instance, in imaging applications with Fourier component measurements, such as VLBI and MRI, since the images are approximately band-limited, the signal sampling rate in the Fourier domain (e.g., $\kappa$-space) can be reduced to twice the maximum frequency~\cite{shannon1949communication}. This observation can be used to define metrics that characterize sampling quality independent of image reconstruction techniques~\cite{palumbo2019metrics}. Modern advances in the theory of compressed sensing demonstrate that, when the underlying image is sparse, certain sampling schemes can achieve the same quality reconstruction with an even smaller number of samples~\cite{compressedsensing, candes2008introduction}. However, both Nyquist sampling and compressed sensing typically consider only linear forward models and assume no noise or only {\it i.i.d.} Gaussian noise, which is unrealistic for many real-world problems.

\subsubsection{Learned Strategies}
Recently, approaches have been proposed to explore the joint optimization of a sensing strategy and image reconstruction method using deep neural networks.
This idea has been used to design the color multiplexing pattern of camera sensors~\cite{chakrabarti2016learning}, the LED illumination pattern in Fourier ptychography~\cite{kellman2019data}, the optical parameters of a camera lens~\cite{EndToEndCam2018}, and the exposure pattern in video compressed sensing~\cite{yoshida2018joint}. However, all these approaches yield only a single deterministic sensing pattern that ignores the possibility of multiple, equally good sensing strategies. In a recent study of ``compressed sensing" MRI, the sensing sampling strategy in the $\kappa$-space is modeled as Bernoulli distributions, so that multiple sensing patterns are explored simultaneously~\cite{bahadir2019adaptive}. This formulation helps characterize which sampling frequencies are important for different anatomy. 
However, none of these prior approaches address complicated non-Gaussian noise, or explicitly model the correlations among measurements.

\section{Proposed Method} \label{sec:method}
We propose a learning method that jointly optimizes the sensor-sampling distribution and the image reconstruction method for forward models that contain correlated measurements and complicated noise. In Sec.~\ref{subsec:jointopt}, we discuss the problem setup. In Sec.~\ref{subsec:ising} we introduce the sensor-sampling distribution, formulated as an Ising model. In Sec.~\ref{subsec:unet} we discuss the image reconstruction sub-network architecture.

\subsection{Joint Sensing and Imaging Optimization} \label{subsec:jointopt}

We formulate the joint sensing and imaging optimization as a deep autoencoder. This autoencoder is trained to identify an efficient image representation that is then used to make limited observations, from which an image can be recovered. The encoder represents the sensing strategy and the decoder represents the image reconstruction method. Both the encoder and decoder are formulated as deep neural networks in this framework. However, to ensure the sensing system design obeys physical constraints, the encoder samples from all possible measurements generated by the physics simulator.

\begin{figure}[!t]
\centering
\setlength{\fboxrule}{0pt}
\framebox[\columnwidth]{\includegraphics[width=1.0\columnwidth]{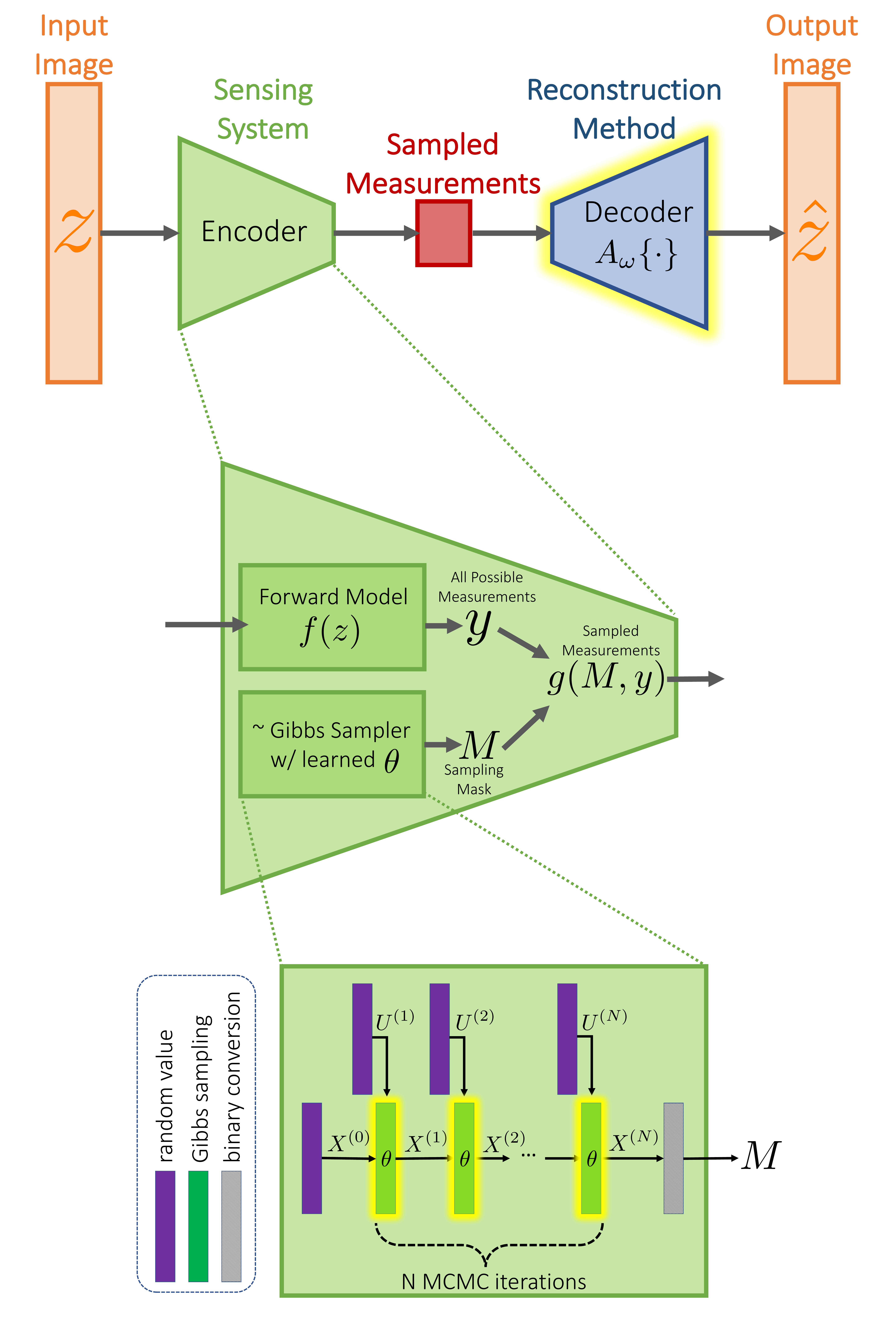}}
\caption[LoF entry]{The architecture of the proposed joint sensing and imaging optimization problem. 
We draw a connection between optimizing a computational imaging system and learning the parameters of an autoencoder:
the encoder consists of a known physics-based forward model that generates all possible measurements and a sensor-sampling block that chooses among those measurements; 
the decoder drives image reconstruction from measurements. We combine the proposed decoder and physically-motivated encoder into a single, differentiable neural network. By maximizing the similarity between the input and output images, the sensor-sampling strategy and image reconstruction are jointly optimized. The trainable blocks, sensor-sampling distribution and reconstruction method, are highlighted in yellow, where the corresponding trainable parameters are $\theta$ and $\omega$.

\hspace{0.1in}The architecture of the sampling sub-network is inspired by the Gibbs sampling procedure. Each Gibbs sampling layer acts as an MCMC iteration: it takes samples $X^{(i-1)}$ from the previous layer and generates more likely samples $X^{(i)}$ based on the states' joint distributions (parameterized by $\theta$). The input to the sub-network is a set of uniform random numbers $U^{(i)}$ and an initial random set of states, $X^{(0)}$, $\in[-1,1]$. After $N$ Gibbs layers, the states are converted to approximately binary by using a sigmoid function with large slope. Mathematical representations of the successive Gibbs sampling layers and the binary conversion layer are shown in Eq.~\eqref{eq:binary1} and Eq.~\eqref{eq:binary2}, respectively.
}
\label{fig:autoencoder}
\end{figure}


This optimization problem can be written as
\begin{equation} \label{eq:loupe}
    \begin{split}
        \arg \min_{\theta, \omega}\quad &\mathbb{E}_{M \sim p_{\theta}(M)} \left[\sum_{j=1}^{n} s(\hat{z}_j, z_j) + \lambda \mathcal{R}(M) \right],\\
        s.t. \quad &\hat{z}_j = A_{\omega}\{g(M, f(z_j))\},
    \end{split}
\end{equation}
where $M$ is a sensor-sampling pattern, whose distribution is $p_{\theta}(\cdot)$ parameterized by $\theta$, $s(\hat{z}_j, z_j)$ is a function defining the similarity between the reconstruction result $\hat{z}_j$ and the true image $z_j$ in the $n$ image training set, $\mathcal{R}(\cdot)$ is the regularization on the sensing samples, $A_{\omega}\{\cdot\}$ is the image reconstruction neural network parameterized by $\omega$, and $g(\cdot,\cdot)$ is the sensing function that produces the observed measurements. 
In this loss function, the first term quantifies the image reconstruction performance, while the second term quantifies desired properties of the sensor-sampling pattern (e.g., sparsity; see Sec.~\ref{subsec:modelreg}). These two terms are balanced by the hyper-parameter $\lambda$. The sampling pattern $M \in [0,1]^K$ is a binary vector that indicates which of the $K$ sensor sources is selected or excluded from observations. This is equivalent to selecting elements from the forward model's vector of measurements, i.e.
\begin{equation} \label{eq:sampling}
g(M, f(z_j)) = S(M) \odot f(z_j),
\end{equation}
where $S(\cdot)$ is the binary selection function based on the mask $M$, whose dimensionality is the same as the forward model $f(\cdot)$. This formulation is a generalization of~\cite{bahadir2019adaptive}, which used a similar optimization designed specifically for MRI acquisition.

Based on Eq.~\eqref{eq:loupe} and Eq.~\eqref{eq:sampling}, we build a joint sensing and imaging autoencoder, as shown in Fig.~\ref{fig:autoencoder}. The trainable blocks in our neural network are the sensor-sampling distribution $p_{\theta}(\cdot)$ (from which the mask $M$ is sampled) and the reconstruction network $A_{\omega}(\cdot)$, where $\theta$ and $\omega$ are the trainable network parameters. In the following sections we describe how to build Eq.~\eqref{eq:loupe} as a single, differentiable neural network and solve for $\theta$ and $\omega$ simultaneously.




\subsection{Physics-Constrained Encoder} \label{subsec:ising}

The goal of the encoder in Fig.~\ref{fig:autoencoder} is to select measurements from the physics-based forward model to pass on to the decoder. The forward model and procedure for masking measurements are both well defined; the challenge is to design a network architecture that can sample a mask from a parameterized distribution. In order to capture the correlations among observations, we model the proposed sensor-sampling distribution $p_\theta(\cdot)$ in Eq.~(\ref{eq:loupe}) as a fully-connected, Ising model (binary Markov random field).
In this section we explain the Ising model sensor-sampling distribution and describe how to sample from this model using a differentiable network architecture.

\subsubsection{Ising model}
In an Ising model, the state of an $n$-element system is represented by a set of binary random variables, $X=(x_1, \cdots, x_n), x_j \in \{+1, -1\}, \forall j \in [1, n]$. The total energy of a fully-connected Ising model is given by the Hamiltonian function,
\begin{equation}
    H_{\theta}(X) = - \sum_j \theta_{jj} x_j - \sum_{j < k} \theta_{jk} x_j x_k,
\end{equation}
where $\theta_{jk}$ are the Ising model parameters. The first term models the energy of single elements and the second term models the potential from the interactions between pairs of elements. The state's probability can be derived according to the Boltzmann distribution
\begin{equation}
    p_{\theta}(X) = \frac{\exp[-H_{\theta}(X)]}{Z_{\theta}},
\end{equation}
where $Z_{\theta}$ is the partition function,
\begin{equation}
    Z_{\theta} = \sum_{X} \exp[-H_{\theta}(X)].
\end{equation}
The lower the total energy of a state, the higher the probability it will occur. 

We use a fully-connected Ising model to model the presence (+1) or absence (-1) of a particular sensor, and capture correlations among sensors. Parameter $\theta_{jj}$ characterizes how important it is to include the $j$-th sensor, and $\theta_{jk}$ characterizes how the $j$-th and the $k$-th sensor are positively or negatively correlated: if $\theta_{jk} > 0$, observing the $j$-th and the $k$-th sensor are favored or disfavored together; while if $\theta_{jk} < 0$, the $j$-th sensor is disfavored when the $k$-th is selected, or visa versa. 
Although an Ising model only explicitly defines pairwise correlations, higher order correlations among sensors are implicitly captured by the model. These higher order correlations can be partially extracted by interpreting the Ising model as an affinity graph and performing clique analysis on the pruned edges (see an example in Sec.~\ref{subsubsec:noise}).

Solving the optimization problem in Eq.~\ref{eq:loupe} requires sampling the Ising model according to parameters $\theta = \{\theta_{jk} | \forall j,k \in [1, n]$\}. However, unlike simple distributions such as Bernoulli, Uniform, or Gaussian, it is difficult to sample an Ising model directly, because the binary variables are correlated and the number of possible states grows exponentially with the dimension of elements. To build a computationally efficient and fully differentiable sampling network, we propose a Markov chain Monte Carlo (MCMC) inspired method.



\subsubsection{Differentiable Gibbs Sampling Sub-Network} \label{subsubsec:mcmc}
We design a sub-network to perform successive MCMC sampling iterations of an Ising model (parameterized by $\theta$), which enables sampling masks from $p_{\theta}(M)$ and facilitates computing Eq.~\eqref{eq:loupe}.

MCMC methods have been widely used for approximating complicated statistical distributions. In a commonly used MCMC method, Gibbs sampling, one samples from the joint probability distribution by conditionally sampling each variable in turn until convergence. The detailed procedure of Gibbs sampling for a binary state space can be summarized as:
\begin{enumerate}
    \item Randomly draw an initial sample
    \begin{equation}
    X^{(0)} =(x_1^{(0)}, \cdots, x_n^{(0)}),
    \end{equation}
    from an independent Bernoulli distribution.
    \item Compute the next sample $X^{(i)}$ based on $X^{(i-1)}$ with each component $x_j^{(i)}$ sequentially updated using its conditional distribution given the other states 
    \begin{equation}
    p_j^{(i)} = p_{\theta}(x_j^{(i)}=1 | x_1^{(i)}, \cdots, x_{j-1}^{(i)}, x_{j+1}^{(i-1)}, \cdots, x_n^{(i-1)})
    .\end{equation}
    For each $p_j^{(i)}$, draw a random number from a uniform distribution
    \begin{equation}
    u_j^{(i)} \sim U(0, 1).
    \end{equation}
    If $u_j^{(i)} < p_j^{(i)}$, assign $x_j^{(i)}=+1$; otherwise, assign $x_j^{(i)}=-1$.
    \item Repeat step 2 for $N$  iterations.
\end{enumerate}
Gibbs sampling is often used to sample from Ising models, as the conditional probability of an Ising model can also be described as an Ising model:
\begin{equation}
    \begin{split}
        &H_{\theta}(\{x_j\} \in \Lambda_{u} | \{x_l\} \in \Lambda_{k}) \\
        &= - \sum_{\substack{j \\ x_j \in \Lambda_{u}}} (\theta_{jj} + \sum_{\substack{l \\ x_l \in \Lambda_{k}}} \theta_{jl} x_l) x_j - \sum_{\substack{j < k \\ x_j, x_k \in \Lambda_{u}}} \theta_{jk} x_j x_k\\
        &= - \sum_{\substack{j \\ x_j \in \Lambda_{u}}} \theta_{jj}^{\prime}(\{x_l\} \in \Lambda_{k}) x_j - \sum_{\substack{j < k \\ x_j, x_k \in \Lambda_{u}}} \theta_{jk} x_j x_k
    \end{split}
    \label{eq:conditional}
\end{equation}
where $\Lambda_{k}$ and $\Lambda_{u}$ are the known and unknown/to-be-updated subsystems respectively, and $\theta_{jj}^{\prime}$ defines the new Ising model parameter conditioned on the known subsystem. Eq.~\eqref{eq:conditional} indicates the conditional distribution of a single element is  a Bernoulli distribution, making it possible to implement each Gibbs sampling iteration as a neural network layer.

In order to sample a mask $M$ from the distribution $p_{\theta}(M)$, we design a neural network where each dense layer corresponds to an iteration of Gibbs sampling for an Ising model. Each Gibbs sampling layer can be mathematically described as a function of $\theta$:
\begin{equation} \label{eq:mcmclayer}
    x_{j}^{(i)} = \text{sgn} \left(\sigma (2 (\sum_{k<j} \theta_{jk} x_k^{(i)} + \theta_{jj} + \sum_{k>j} \theta_{jk} x_k^{(i-1)})) - u_{j}^{(i)}\right),
\end{equation}
where $\text{sgn}(\cdot)$ is a sign step function and $\sigma(\cdot)$ is a sigmoid function,
\begin{equation} 
    \text{sgn}(y) = 
    \begin{cases}
    -1 & y \leq 0 \\
    +1 & y > 0
    \end{cases}, \quad
    \sigma(y) = \frac{1}{1+\exp{(-y)}}.
\end{equation}
The initial sample $X^{(0)}$ is initialized using uniform random values, $x_j^{(0)} = \text{sgn}(0.5 - u_j^{(0)})$. Eq.~\eqref{eq:mcmclayer} defines a layered Ising sampling function, which converges to a sample from the Ising model after a sufficient number $N$ of iterations. 
However, the sign function in Eq.~\eqref{eq:mcmclayer} has no gradient (zero or undefined), rendering back-propagation strategies impossible. To address this, we replace the sign function with a hyperbolic tangent function
\begin{equation} \label{eq:binary1}
    x_{j}^{(i)} \hspace{-.07in} = \text{tanh}_{s_1}\hspace{-.07in}\left(\sigma(2(\sum_{k<j} \theta_{jk} x_k^{(i)} + \theta_{jj} + \sum_{k>j} \theta_{jk} x_k^{(i-1)})) - u_{j}^{(i)}\right)\hspace{-.05in},
\end{equation}
where $\text{tanh}_{s_1}(a) = \text{tanh}(s_1 a)$.
Using a moderate $s_1$ slope enables us to approximately binarize the states to $[-1,1]$ between successive Gibbs layers, while still allowing gradients to propagate through the network.

To obtain the sampling mask, $M$, we apply a strong sigmoid function to the final sampling layer of the Ising model to convert the states from approximately $[-1,1]$ to nearly binary states $[0,1]$:
\begin{equation} \label{eq:binary2}
    M = \sigma_{s_2}(X^{(N)}) = \sigma({s_2} X^{(N)}),
\end{equation}
where $s_2$ is a slope of the sigmoid function that should be large enough to approximate the Heaviside function well. 


For notational simplicity, we represent this sampling procedure as a nested function,
\begin{equation}
    \begin{split}
    M &\sim p_{\theta}(M) \\
    & = \sigma_{s_2}(X^{(N)}) \\
    &= \sigma_{s_2}(G_{\theta}(U^{(0)}, \cdots, U^{(N)})),
    \end{split}
\end{equation}
with Ising model coefficient $\theta$ as its parameters, uniformly distributed random numbers, $\{U^{(0)}, \cdots, U^{(N)}\}$ as inputs, and the $[0,1]$ binary samples in $M$ as outputs. The architecture of the Gibbs sampling neural network is shown at the bottom of Fig.~\ref{fig:autoencoder}.

\subsubsection{Ising Model Regularization}
\label{subsec:modelreg}
To encourage sparse samples and promote sample diversity (in order to explore multiple observational strategies), we define the regularization on Ising samples ($M \sim p_{\theta}(M)$) as consisting of two terms:
\begin{equation} \label{eq:regularizer}
    \mathcal{R}(M) = \lambda_1 \|M\|_1 - \lambda_2 H_{\theta}(X^{(N)}),
\end{equation}
where the first term is a $\ell_1$ regularization on the number of participating sensors (sparsity loss), and the second term is the negative Hamiltonian of the Ising samples (diversity loss). The Hamiltonian of an Ising model is a good approximation of the distribution entropy as it only differs in its log partition function,
\begin{equation} \label{eq:entropy}
    \begin{split}
        S_{\theta} &= -\sum_{X} p_{\theta}(X) \log p_{\theta}(X)\\
        &= -\sum_{X} p_{\theta}(X) [-H_{\theta}(X) - \log Z_{\theta}]\\
        &= \mathbb{E}_{X \sim p_{\theta}(X)} [H_{\theta}(X)] + \log Z_{\theta}.
    \end{split}
\end{equation}
Therefore, maximizing the Hamiltonian typically increases the system entropy, thus diversifying the samples from our Ising model. Substituting Eq.~(\ref{eq:regularizer}) into Eq.~(\ref{eq:loupe}) and approximating the expectation using the empirical average via MCMC samples, we derive a stochastic formulation of the joint optimization problem under Ising sampling,
\begin{equation} \label{eq:loupe2}
    \begin{split}
        \arg \min_{\theta, \omega}\quad &\sum_{j=1}^{n} s(\hat{z}_j, z_j) + \lambda_1 \|M_j\|_1 - \lambda_2 H_{\theta}(X_j^{(N)}),\\
        s.t. \quad &\hat{z}_j = A_{\omega}\{S(M) \odot f(z_j)\}, \\
        & M_j = \sigma_s(X_j^{(N)}), \\
        & X_j^{(N)} = G_{\theta}(U_j^{(0)}, \cdots, U_j^{(N)}), \\
        & U_j^{(0)}, \cdots, U_j^{(N)} \sim U(0, 1).
    \end{split}
\end{equation}

\subsection{Image Reconstruction Decoder} \label{subsec:unet}
The image reconstruction decoder sub-network is not restricted to any particular neural architecture, but should have enough model capacity to be able to reconstruct with multiple sensing scenarios simultaneously. For imaging problems with a complicated forward model and noise, physics-based neural networks (PbNN)~\cite{kellman2019data} and unrolled neural networks~\cite{diamond2017unrolled, sun2016deep} have proven successful. However, with sufficient training data, general deep learning frameworks such as the U-Net and its variants~\cite{ronneberger2015u, lee2017deep} often perform sufficiently well. The U-Net combines local and global information during image reconstruction by incorporating hierarchical down-sampling and up-sampling blocks and multiple skip connections in its architecture.
In our VLBI case study, two variants of the U-Net are used for reconstruction. We discuss the details of their architectures in Sec.~\ref{subsubsec:nn}.

\section{Case Study: VLBI Array Design} \label{sec:result}

Very-Long-Baseline-Interferometry (VLBI) is a radio interferometry technique for high-resolution astronomical imaging. In VLBI, telescopes are linked to form an imperfect, virtual telescope with a maximum resolution defined by the longest baseline between participating telescopes~\cite{thompson1986interferometry}. Recently, this technique was used by Event Horizon Telescope (EHT) to capture the first image of a black hole, the image of M87*~\cite{akiyama2019first}, by joining telescopes (observing at 230 GHz) from across the globe. Building on this recent success, the EHT is continuing to improve its telescope array by adding new sites, in order to improve M87*'s image quality as well as image additional astronomical sources, such as the Milky Way's black hole, Sagittarius A*.

It is important to carefully study which new sites should be added to the next-generation EHT to best improve future results. Adding a new telescope to the existing EHT VLBI array is extremely expensive: equipping a pre-built telescope to work as part of the array costs over two million dollars, and the construction of a new telescope can easily cost over 10 million dollars\cite{blackburn2019studying}. Therefore, limited budgets require selecting only a handful of new sites that can be added to the future array. Since only a select number of ground sites are suitable for constructing telescopes that can observe at 230 GHz (e.g., due to altitude and weather), the problem of choosing telescope locations for the next-generation EHT reduces to a site selection problem. Studying the impact of a telescope on imaging quality is not only important for planning an EHT expansion but also becomes a powerful tool during an observing campaign; when coordinating future observations it is essential to evaluate the impact of losing a particular telescope -- should costly observations proceed if weather is poor or there are instrumental problems at a particular site?



\begin{figure}[!t]
\centering
\setlength{\fboxrule}{0pt}
\framebox[\columnwidth]{\includegraphics[width=1.0\columnwidth]{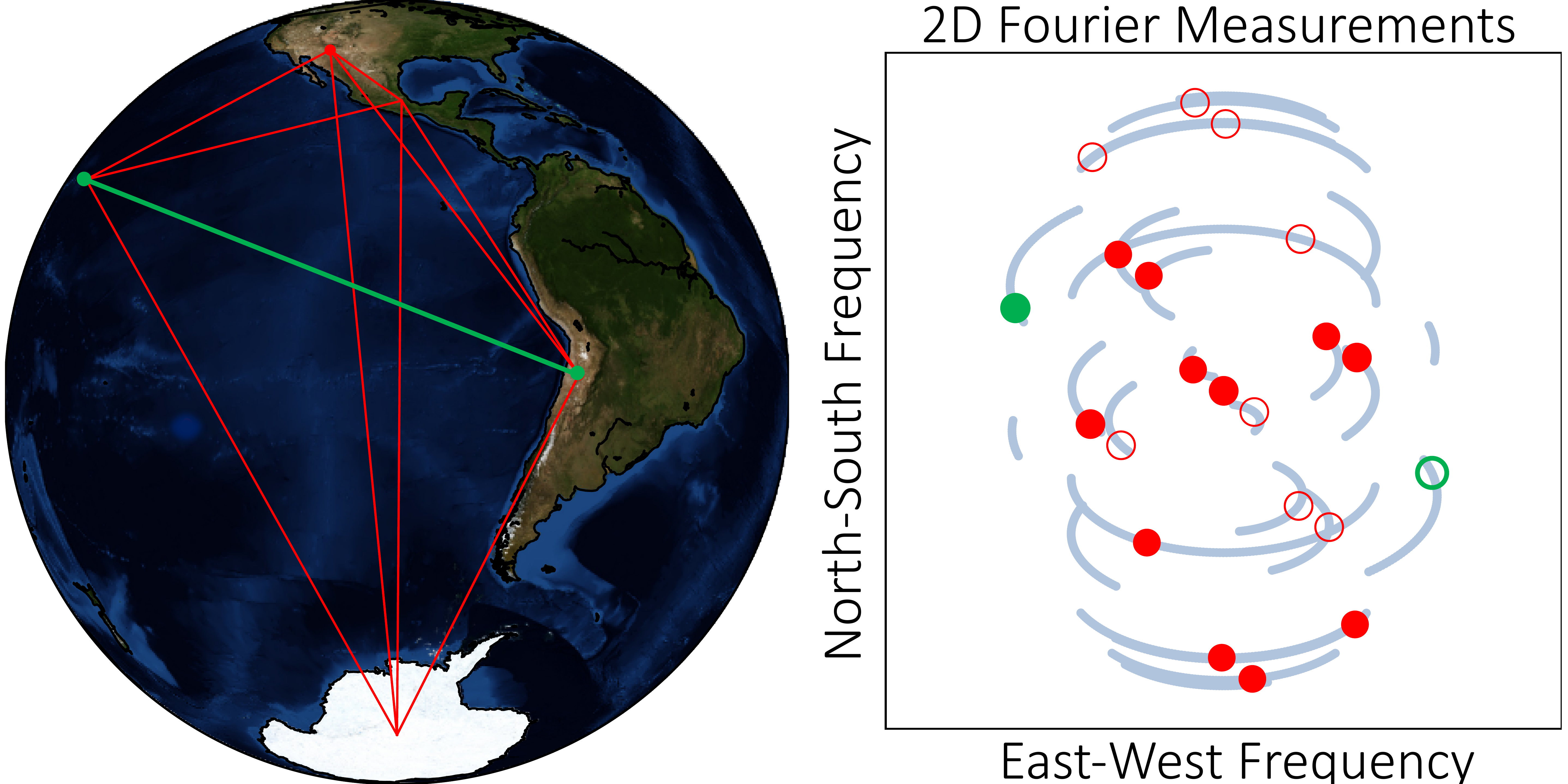}}
\caption{VLBI baselines and measurements. The left figure indicates the baselines between telescopes in the EHT array at a particular time, where the Earth is being viewed from the direction of Sgr A*. The right figure shows the corresponding 2D spatial-frequency coverage of measurements collected by the telescopes. Each dot represents a visibility measurement, where the 2D spatial-frequency of the visibility is related to the projected baselines between two telescopes at a given time. An example baseline and its Fourier measurement (two symmetric dots due to the constraint the the image is real-valued) are marked in green. As Earth rotates, the projected baseline vectors change, resulting in a more complete frequency coverage (marked with light blue) collected over a night.}
\label{fig:uvcov}
\end{figure}

We optimize the design of a VLBI array by solving a site selection problem.  The goal is to decide which sites from a given list we should build into the array, and which sites can be excluded without substantial loss to the reconstructed image. 
Although many factors are ultimately important for site selection (e.g., construction cost, weather stability), in this paper we focus solely on the location of sites, as the array geometry usually has the largest influence over resulting image quality. As explained in Sec.~\ref{subsec:probset}, VLBI measurements are produced from data collected at multiple telescopes, so it is important to consider how telescopes interact with one another. 
We capture the telescope correlations through the fully-connected binary Ising model, and optimize the telescope sampling distribution jointly with the image reconstruction method using the proposed approach.


\subsection{Problem Setup}
\label{subsec:probset}
In radio interferometry (e.g., VLBI), data simultaneously collected from a pair of telescopes is used to produce a single measurement. Each measurement corresponds to a two-dimensional Fourier component of the image, with the frequency related to the baseline between the two participating telescopes, projected in the direction of the observed astronomical source. This measurement is typically referred to as a {\it visibility}. For a $K$ telescope array, there are ${K \choose 2}$ possible visibility measurements at a single time. Due to Earth's rotation, the projected baselines change over time, causing sparse elliptical measurements in the spatial-frequency domain to be collected over the course of a night (see Fig.~\ref{fig:uvcov}). 

The measurement equation for each visibility, $V$, is given by 
\begin{equation}
    V_{p,q}^t = g_p^t g_q^t \exp\left[-i(\phi_p^t - \phi_q^t)\right] F_{p,q}^t z + n_{p,q}^t,
\end{equation}
for telescope pair $(p,q)$. $F_{p,q}^t z$ extracts a Fourier component from the image $z$ corresponding to the baseline between telescopes $p$ and $q$ at time $t$. Each measurement experiences time-dependent telescope-based gain, $g$, and phase error, $\phi$, and baseline-based thermal noise, $n$, where $n_{p, q}^{t} \sim \mathcal{N}(0, \nu_{p, q}^2)$. 
The standard deviation of the thermal noise depends on each telescope's ``system equivalent flux density" (SEFD):
\begin{equation}
    \nu_{p, q} \propto \sqrt{SEFD_{p} \times SEFD_{q}},
\end{equation}
%
where a higher SEFD indicates the measurements using that telescope contain more thermal noise (i.e., lower SNR)~\cite{thompson1986interferometry}.
When atmospheric turbulence is not considered, $g=1$ and $\phi=0$. Although the absolute gains $g$ can often be sufficiently calibrated, phase errors due to atmospheric turbulence cannot be calibrated {\it a priori} since $\phi$ varies rapidly and is uniformly sampled from $[0,2\pi)$\cite{thompson1986interferometry}. Thus, in the case of atmospheric error, recovering an image amounts to solving a phase retrieval problem.

In the case of atmospheric noise, data products called closure quantities can be used to constrain the image reconstruction~\cite{chael2018interferometric}. Closure phase is obtained by multiplying three visibilities from telescopes in a closed loop:
\begin{equation}
    C_{p,q,b}^t = \angle \left( V_{p,q}^t V_{q,b}^t V_{b,p}^t\right)
\end{equation}
Since the atmospheric errors are site dependent, $C_{p,q,b}$ is robust to any phase errors $\phi$. Data products called closure amplitudes are robust to gain errors~\cite{chael2018interferometric}. Since amplitudes can often be much better calibrated than phases, in this paper we do not consider closure amplitudes but make use of visibility amplitudes, $|V_{p,q}^t|$.


\begin{table}[!t]
\renewcommand{\arraystretch}{1.3}
\caption{Telescope sites in "EHT+" array. }
\centering
\begin{tabular}{c||c|c}
\hline
Sites & Location & SEFD\\
\hline\hline
PV$^{\star}$ & Pico Veleta, Spain & 1400 \\
\hline
PDB & Plateau de Bure, France & 1500 \\
\hline
ALMA$^{\star}$ & Atacama Desert, Chile & 90\\
\hline
APEX$^{\star}$ & Atacama Desert, Chile & 3500\\
\hline
LMT$^{\star}$ & Sierra Negra, Mexico & 600\\
\hline
SMT$^{\star}$ & Mt. Graham, Arizona & 5000\\
\hline
SPT$^{\star}$ & South Pole & 5000\\
\hline
OVRO & Owens Valley, California & 10000\\
\hline
JCMT$^{\star}$ & Maunakea, Hawai'i & 6000\\
\hline
SMA$^{\star}$ & Maunakea, Hawai'i & 4900\\
\hline
KP & Tucson, Arizona & 10000\\
\hline
GLT & Thule Air Base, Greenland & 10000\\
\hline
\end{tabular}
\label{tab:sites}
\end{table}


\subsection{Implementation Details} \label{subsec:implementation}

\subsubsection{Telescope Arrays}

\begin{figure}[!t]
\centering
\setlength{\fboxrule}{0pt}
\framebox[\columnwidth]{\includegraphics[width=1.0\columnwidth]{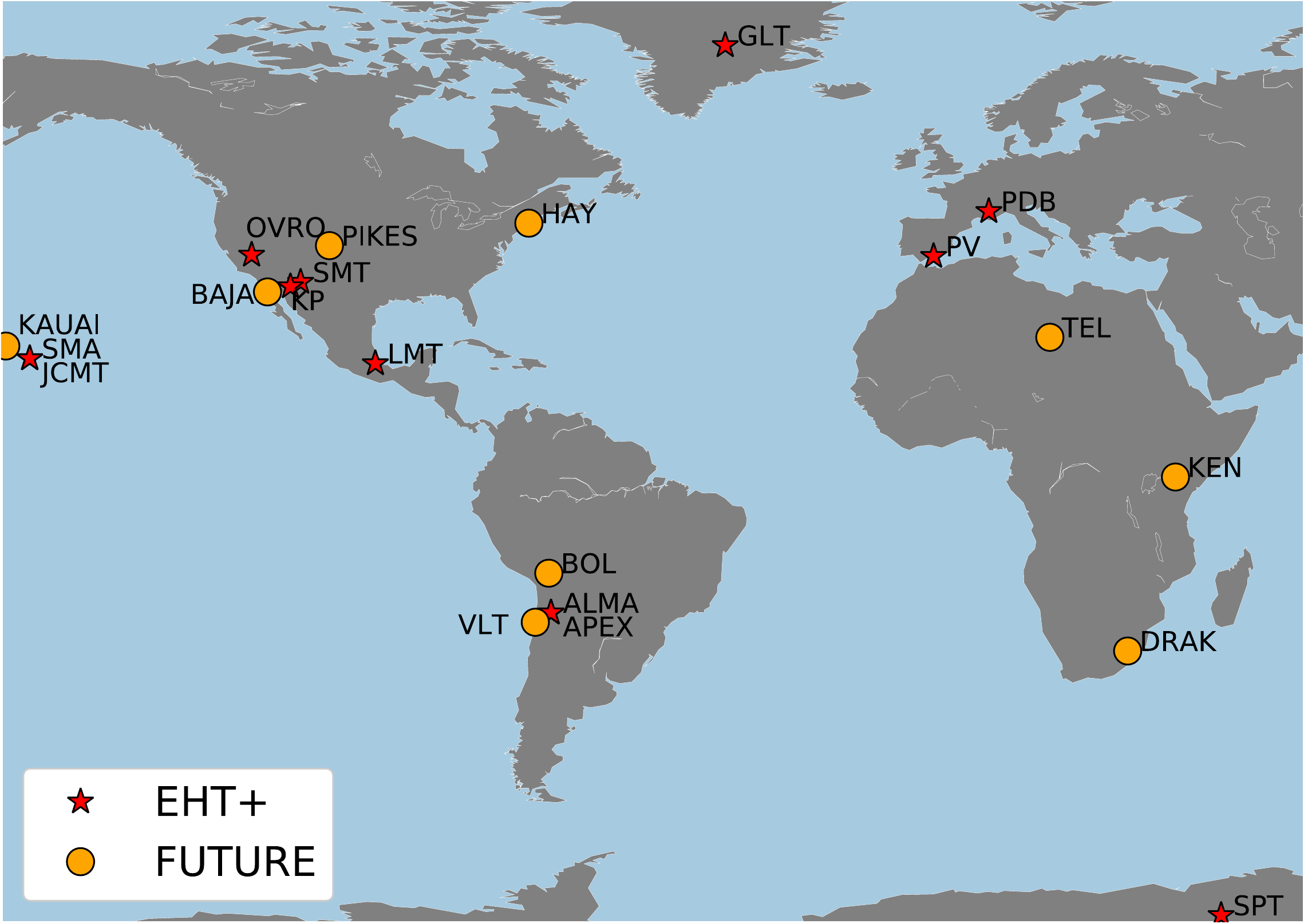}}
\caption{Site map of potential future EHT telescope locations. Twelve sites (``EHT+") marked with blue stars are existing telescopes currently participating in or planning to join the EHT. The other nine sites (``FUTURE"), marked with orange dots, are potential locations where new telescopes could be added. ``FUTURE" sites are selected as locations that can observe at the necessary 230 GHz (1.3 mm wavelength) observed by the EHT.}
\label{fig:map}
\end{figure}

In this paper, we consider two arrays of radio telescopes: ``EHT+'' and ``FUTURE" (Fig.~\ref{fig:map}). ``EHT+" includes twelve telescopes sites, as listed in Table~\ref{tab:sites}. Eight of these sites (marked with a star) were used by the EHT in 2017~\cite{akiyama2019first}, while the other four sites host existing radio telescopes that plan to eventually join the EHT. Each telescope's SEFD (i.e., noise coefficient) is reported in the table. The ``FUTURE" array consists of nine additional sites that do not currently host a suitable telescope, but have sufficient atmospheric conditions for a constructed telescope to successfully observe at 230 GHz with the EHT.\footnote{Our analysis conservatively assumes each new telescope in the ``FUTURE" array has an SEFD of 10000.}

\subsubsection{Ising Model Sampler}
The VLBI forward model $f(\cdot)$ produces as a vector of complex visibilities or visibility amplitudes and closure phases. The mask, $M$, is a binary vector representing whether each telescope is selected for observation. We define the masking function for each VLBI measurement in $f(\cdot)$ as $S_{p,q}(M) = M_p M_q$ for a visibility $V_{p,q}$ or visibility amplitude $|V_{p,q}|$, and $S_{p,q,b}(M) = M_p M_q M_b$ for a closure phase $C_{p,q,b}$. The mask is generated from the Ising model sampler in Section~\ref{subsec:ising}, which is built using five MCMC layers. The sampling order of telescopes for each MCMC layer is randomly shuffled for each training trial. In Section~\ref{subsec:experiment}, we run the joint optimization five times for each experiment to approximate the mean and standard deviation of the learned Ising model parameters. The slopes for binary conversion layers (Eq.~\ref{eq:binary1} and Eq.~\ref{eq:binary2}) are empirically set to $s_1 = 3$ and $s_2 = 10$.

\subsubsection{Image Reconstruction Architectures} \label{subsubsec:nn}

\begin{figure*}[!t]
\centering
\setlength{\fboxrule}{0pt}
\framebox[\textwidth]{\includegraphics[width=1.0\textwidth]{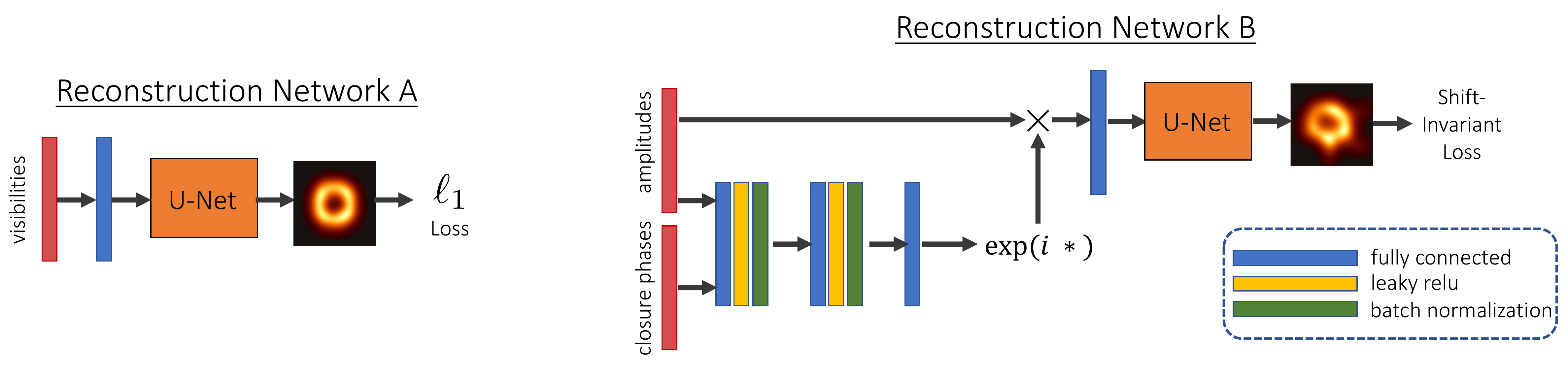}}
\caption{VLBI reconstruction sub-network architectures. Reconstruction Network A is designed to recover an image from complex visibility measurements that experience, at most, baseline-varying Gaussian thermal noise. Reconstruction Network B is designed to recover an image when measurements experience corrupting atmospheric phase error in addition to thermal noise; since the absolute phase information in the complex visibilities are lost in this case, the network takes visibility amplitudes and closure phases as input (rather than complex visibilities) and uses a shift-invariant loss (rather than a simpler $\ell_1$ loss). 
For each network variant we show a single representative test reconstruction obtained from measurements, which corresponds to the ring truth image shown in Fig.~\ref{fig:blur_recon}.
}
\label{fig:reconnet}
\end{figure*}

Two different neural network architectures (Fig.~\ref{fig:reconnet}) are used for the image reconstructions (Section~\ref{subsec:unet}), with input corresponding to either complex visibilities or visibility amplitudes and closure phases, as shown in Fig.~\ref{fig:reconnet}. Both networks output a $32\times32$ pixel image.

In the case of no atmospheric turbulence, Reconstruction network A is used (Fig.~\ref{fig:reconnet}). The input to this network are complex visibilities. The network architecture consists of one fully connected layer and a U-Net (four down-size layers + four up-size layers). Ideally, the fully connected layer transforms the Fourier measurements from the frequency domain back to the image domain. Then, the U-Net removes aliasing effects caused by missing data and noise. Since the image's phase information is preserved when there is no atmospheric turbulence, we can define the image similarity loss, $s(\cdot, \cdot)$ as an $\ell_1$ norm of the difference between the reconstructed and blurred original image:
\begin{equation}
    s(\hat{z}_j, z_j) = \| \hat{z}_j - \mathcal{K}(res) * z_j\|_1,
\end{equation}
where $\mathcal{K}$ is a Gaussian kernel that defines our desired reconstruction resolution. The nominal resolution of our telescope array ($\approx25$ $\mu$as full-width-half-max (FWHM)) is found using the longest baseline. Image reconstructions that recover structure smaller than the nominal resolution are super-resolving the target.

In the case of atmospheric turbulence, reconstruction network B is used (Fig.~\ref{fig:reconnet}). The input is constructed from visibility amplitudes and closure phases (rather than complex visibilities). The reconstruction network architecture is designed through physical intuition in order to handle the more challenging phase-retrieval problem. First, the visibility amplitudes and closure phases are sent to three general dense layers whose purpose is to disentangle the corrupted visibility phases. Second, the disentangled phases are combined with the visibility amplitudes to produce modified complex visiblities. These modified complex visibilities are then passed to the same architecture used above (Dense + U-Net).  
Since the absolute phase is lost in the presence of atmospheric noise, we use a shift-invariant loss that represents the negative inner product between the true and reconstructed images: 
\begin{equation} \label{eq:shiftinvariant}
    s(\hat{z}_j, z_j) = 1 - \frac{\max \{\hat{z}_j * (\mathcal{K}(res) * z_j)\}}{ \|\hat{z}_j\|_2 \|\mathcal{K}(res) * z_j\|_2}.
\end{equation}

\subsubsection{Training} \label{subsubsec:training}
We train our joint optimization problem using images from the ``Fashion-MNIST'' dataset\cite{xiao2017fashion}, and validate on the rest of ``Fashion-MNIST'', classic ``MNIST'', and simple geometric model images meant to resemble a black hole. Each image is $32 \times 32$ pixels and corresponds to a 100 $\mu$-arcsecond field of view (FOV) with a total flux of 1 Jansky (Jy), unless otherwise specified. The training set is augmented by introducing local deformations and random rotations to the original ``Fashion-MNIST'' images. Data augmentation is found to be especially important for training reconstruction network B to avoid overfitting. Each network is trained until convergence with Adam in TensorFlow\cite{kingma2014adam}, requiring between 50-300 epochs with a learning rate of $10^{-3}$. 
VLBI visibity measurements are produced using the eht-imaging Python library\cite{chael2018interferometric}.

\subsection{Experiments}
\label{subsec:experiment}
We demonstrate the proposed optimization framework in a variety of different VLBI scenarios. The purpose of the proposed method is to learn sensor-sampling designs in complex scenarios where intuition is limited. We choose to show some examples in simple scenarios where intuition can be used to confirm that our method provides reasonable results (e.g., Section~\ref{subsubsec:resolution}). However, what is most interesting are the more realistic, complicated cases that cannot be as easily judged from intuition (e.g., Section~\ref{subsubsec:noise}).

\begin{figure}[!t]
\centering
\setlength{\fboxrule}{0pt}
\framebox[\columnwidth]{\includegraphics[width=1.0\columnwidth]{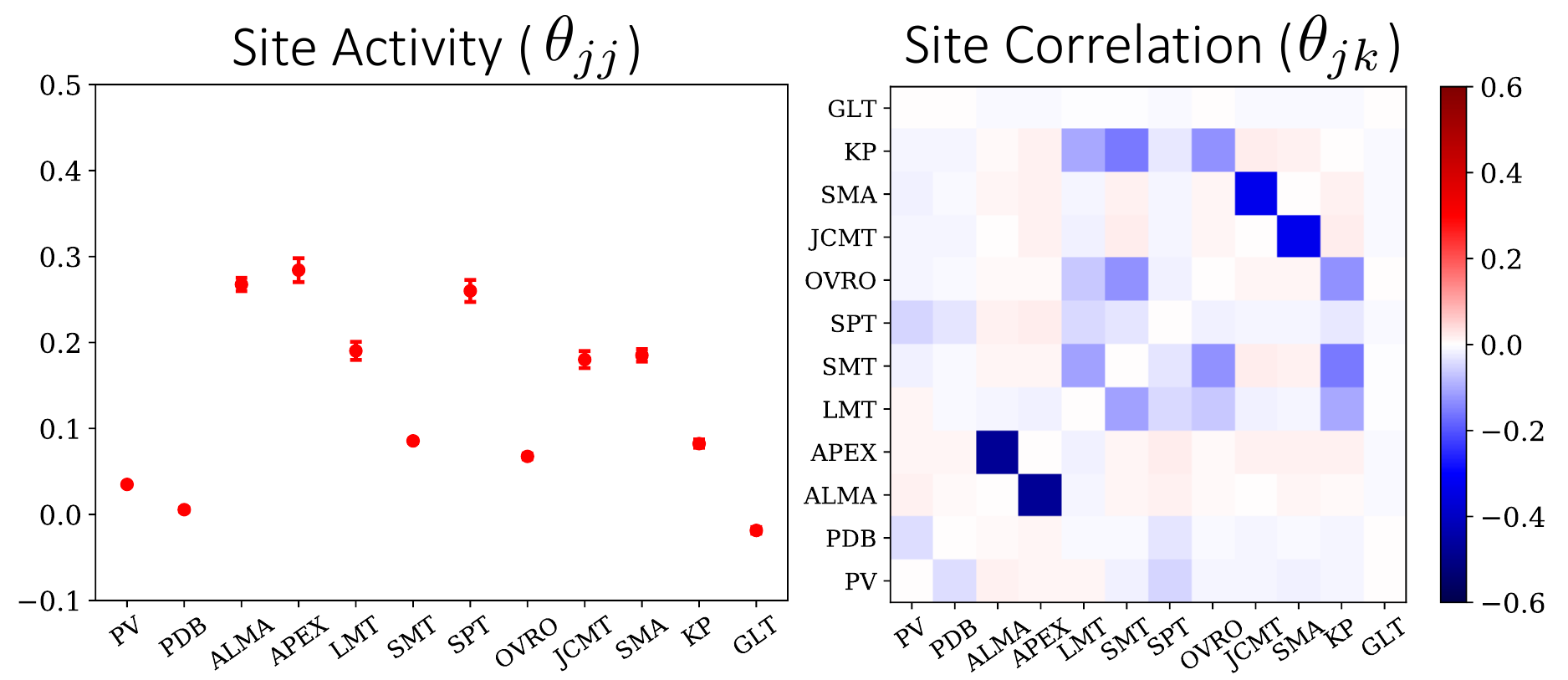}}
\caption{The learned sensor-sampling Ising model parameters. In the VLBI array design problem, $\theta_{jj}$ represents the activity of each site (roughly proportional to how frequently a telescope is sampled), and $\theta_{jk}$ represents the correlation between each pair of telescopes. Representative hyper-parameters (explained in Sec.~\ref{subsec:experiment}) are used in this optimization: the ``EHT+" telescope array, science target Sgr A*, $\lambda_1=\lambda_2=0.005$, $0.75 \times$ the nominal resolution target resolution, and reconstruction network A with complex visibilities. No noise is introduced on the measurements so as to purely investigate the influence of baseline geometry on the recovered distributions.}
\label{fig:ising_param}
\end{figure}

\begin{figure*}[!h]
\centering
\setlength{\fboxrule}{0pt}
\framebox[\columnwidth]{\includegraphics[width=2.0\columnwidth]{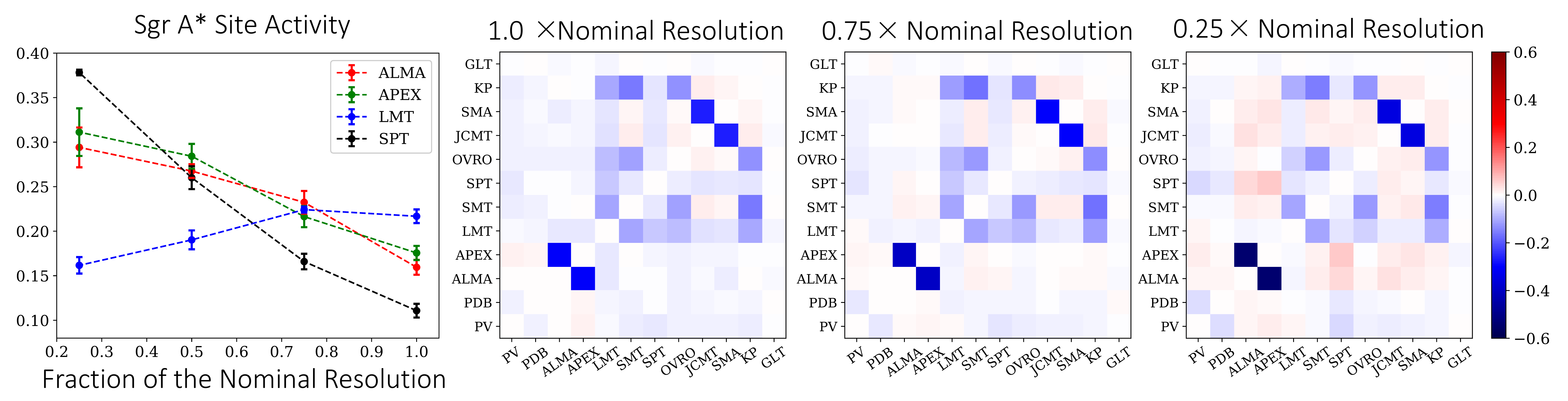}}
\caption{Effect of target reconstruction resolution on array design. Higher resolution requirements (higher resolution corresponds to a smaller fraction of the nominal resolution) increase the learned activity of sites that participate in long baselines, such as SPT, but decrease the activities of sites that participate in short baselines, such as LMT. The long baseline telescope pairs (SPT-ALMA, SPT-APEX, SPT-JCMT and SPT-SMA) become more positively correlated with increased resolution, while zero baseline pairs (APEX-ALMA and JCMT-SMA) become more negatively correlated.}
\label{fig:blurry}

\centering
\setlength{\fboxrule}{0pt}
\framebox[\textwidth]{\includegraphics[width=0.99\textwidth]{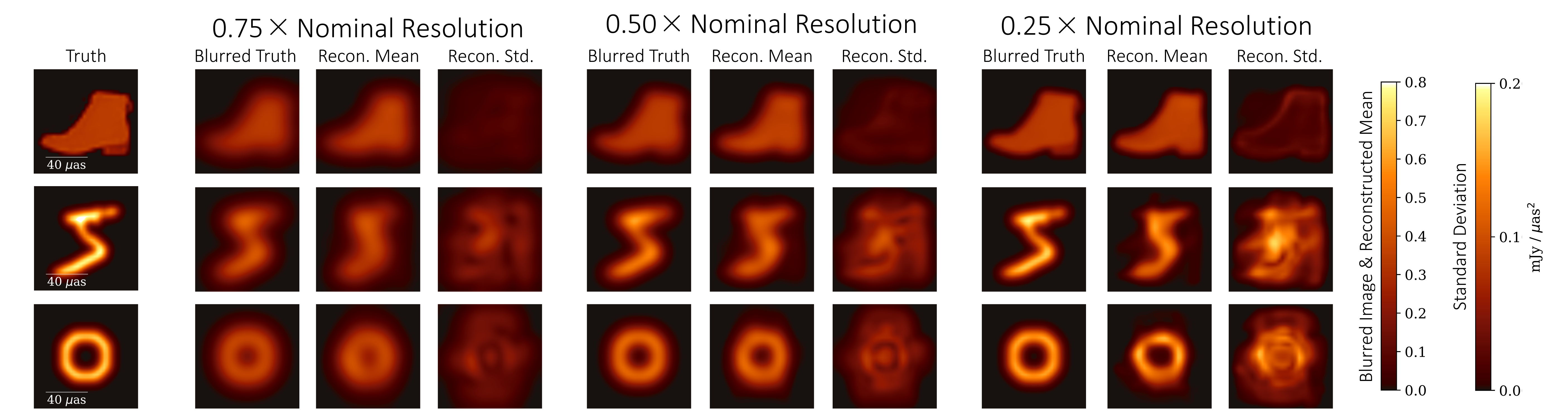}}
\caption{Image reconstructions for different resolution requirements. Reconstruction Network A (refer to Fig.~\ref{fig:reconnet}) was trained as part of the full proposed network shown in Fig.~\ref{fig:autoencoder} at different target resolutions. We show the results of the reconstruction sub-network in three test cases at different target resolutions.
The reconstruction mean and standard deviation are computed based on 1000 samples from the learned sensor-sampling distribution. Each reconstruction sample uses a different mask, $M\sim p_{\theta}(M)$, corresponding to a different set of observing telescopes; the sub-network must be able to handle multiple telescope sampling patterns simultaneously. 
Resolution requirements are specified during training by changing $\mathcal{K}(res)$ in the image similarity loss. The resolution of the array, as defined by the longest baseline (i.e., highest spatial-frequency), is defined as the nominal resolution. Recovering images at a fraction of this resolution (e.g., 0.5 $\times$ and 0.50 $\times$ the nominal array) require super resolving the target. 
}
\label{fig:blur_recon}
\end{figure*}

\begin{figure}[!h]
\centering
\setlength{\fboxrule}{0pt}
\framebox[\columnwidth]{\includegraphics[width=1.0\columnwidth]{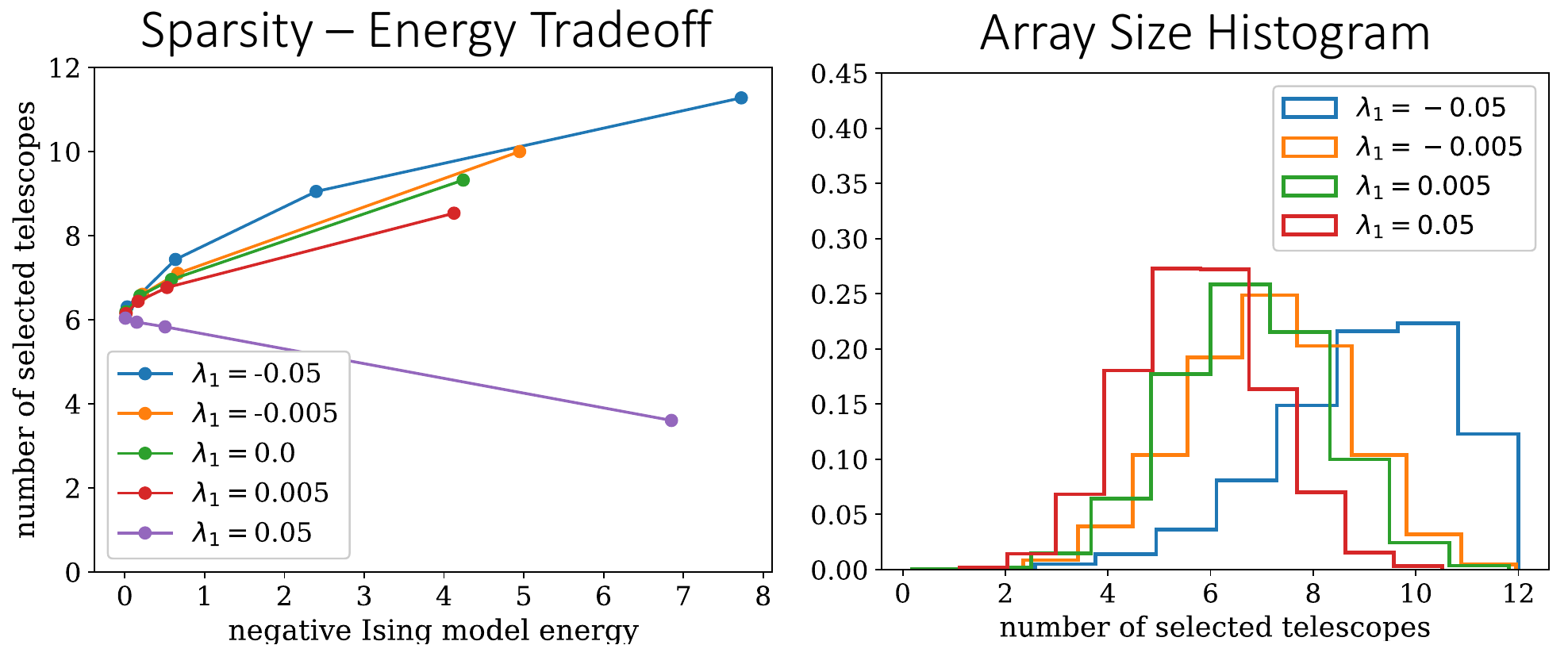}}
\caption{Effect of regularization weights. In the left sparsity-diversity tradeoff figure, we show that (1) sparsity regularization ($\lambda_1$) controls the average number of telescopes selected, and (2) as diversity regularization ($\lambda_2$) is increased, the probability a telescope is selected approaches 50\%, resulting in the average number of telescopes converging to 6 out of 12. The right figure shows the array size (number of telescopes) histogram for a fixed diversity regularization weight ($\lambda_2=0.005$) and changing sparsity regularization weights ($\lambda_1=-0.05, -0.005, 0.005, 0.05$). Larger sparsity regularization drives the sensor-sampling distribution to learn to sample fewer telescopes, as expected.}
\label{fig:gridsearch}
\end{figure}

Fig.~\ref{fig:ising_param} shows a representative Ising model learned from our proposed method. It has been trained using the complex visibility reconstruction network, with representative coefficients $\lambda_1 = \lambda_2 = 0.005$. In this example, we considered the ``EHT+" array with the black hole Sagittarius A$^{\star}$(Sgr A*) as the science target. We define $\mathcal{K}(res)$ in $s(\cdot, \cdot)$ as corresponding to a Gaussian kernel with a FWHM of 0.5 $\times$ the nominal beam. To investigate the influence of baseline geometry on the recovered distributions, we did not include any noise on the measurements.

Note that in the site correlation plot $(\theta_{jk})$, the two pairs of telescopes at same location, ALMA-APEX in Chile and JCMT-SMA in Hawaii, are both highly negatively correlated. This agrees with intuition, as the baselines from any telescope to either co-located telescope will produce identical measurements (when ignoring noise). The parameters, $\theta_{jj}$, represent the activity of each telescope site, and is roughly proportional to how frequently a telescope is sampled. As can be seen, the most important sites in the EHT network for Sgr A* observations are ALMA, APEX, SPT and LMT, which produce measurements covering low to high spatial-frequency. To the contrary, since GLT provides no measurements (it is occluded by Earth since Sgr A* is in the southern hemisphere), it is the least important site and has a negative activity parameter.

\subsubsection{Effect of Sparsity and Diversity Regularization} \label{subsubsec:regularization}
We conduct a grid search on the sparsity and diversity regularization coefficients and explore their influence on the learned optimal sensor-sampling distributions. In our experiments, the diversity regularization coefficient ($\lambda_2$) ranges from $0.001$ to $0.05$, and the sparsity regularization coefficient ($\lambda_1$) ranges from $-0.05$ to $0.05$. Negative sparsity regularization favors observing with more telescopes. Apart from the regularization weights, we use the representative parameters described in the beginning of Section~\ref{subsec:experiment}.

Fig.~\ref{fig:gridsearch} shows the diversity-sparsity relationship recovered from different combinations of regularization weights. The diversity regularizer characterizes the entropy of the Ising model: when the diversity (i.e., the Ising model energy) is large (large $\lambda_2$), all telescope sites have around a $50\%$ chance of being selected or excluded. Therefore, the mean number of selected telescopes gradually converges to six out of twelve with increasing $\lambda_2$. As shown by the array size histogram, increasing the sparsity regularizer ($\lambda_1$) results in a fewer number of sampled telescopes.

\subsubsection{Effect of Reconstruction Resolution} \label{subsubsec:resolution}
We investigate the optimal sensor-sampling for different reconstruction resolution requirements by optimizing the joint sensing and imaging network at four different fractions (1, 0.75, 0.5 and 0.25) of the nominal resolution in $\mathcal{K}(res)$.
Apart from $\mathcal{K}(res)$, we use the representative parameters described in the beginning of Section~\ref{subsec:experiment}. Fig~\ref{fig:blurry} reports the activity of four telescopes (ALMA, APEX, LMT and SPT) that change most with the target resolution, and the learned correlation matrix at 1x, 0.75x and 0.25x the nominal resolution. The correlation matrix at 0.5x the nominal resolution can be found in Fig.~\ref{fig:ising_param}.

As the resolution becomes higher (right to left in the activity plot), LMT becomes increasingly less important, but ALMA, APEX and especially SPT become more important. This agrees with intuition, as SPT baselines measure the highest spatial-frequency information for Sgr A* and LMT is contained in many of the shorter baselines probing low spatial-frequencies. The correlation coefficients between short-baseline pairs (such as APEX-ALMA, JCMT-SMA) become lower at higher resolution, while the correlation coefficients between long-baseline pairs (such as SPT-APEX, SPT-ALMA, SPT-JCMT and SPT-SMA) become higher. This satisfies physical intuition as long baselines are most important for high-resolution reconstructions.

Fig.~\ref{fig:blur_recon} shows the blurred images and reconstruction results at different resolutions. We select one example from each type of validation dataset (Fashion-MNIST, MNIST and geometric models). More precisely, we show the reconstruction mean and standard deviation obtained by taking 1000 telescope samples from each learned model. The higher resolution reconstructions are more challenging as they attempt to super-resolve the image, and as expected they have larger standard deviations. Although the goal of this work is not to develop a new image reconstruction method, we find that the proposed neural network appears to produce promising results for the simplified VLBI reconstruction problem, and is powerful enough to be able to tackle a variety of possible array configurations simultaneously.

\subsubsection{Effect of Noise} \label{subsubsec:noise}
The previous experiments assume no noise in the VLBI measurements. In this section, we compare the optimal telescope sensor-sampling under six different noise assumptions, (1) no noise, (2) equal thermal noise (using the average thermal noise derived from Table~\ref{tab:sites}), (3) site-varying thermal noise (as derived from Table~\ref{tab:sites}), (4) atmospheric phase noise, (5) atmospheric phase noise \& equal thermal noise, and (6) atmospheric phase noise \& site-varying thermal noise. For the latter three cases, the visibility amplitudes and closure phases are used for the image reconstruction. All cases assume a target source of Sgr A* with a constant total flux of 1 Jy using the ``EHT+" array.

\begin{figure}[!t]
\centering
\setlength{\fboxrule}{0pt}
\framebox[\columnwidth]{\includegraphics[width=1\columnwidth]{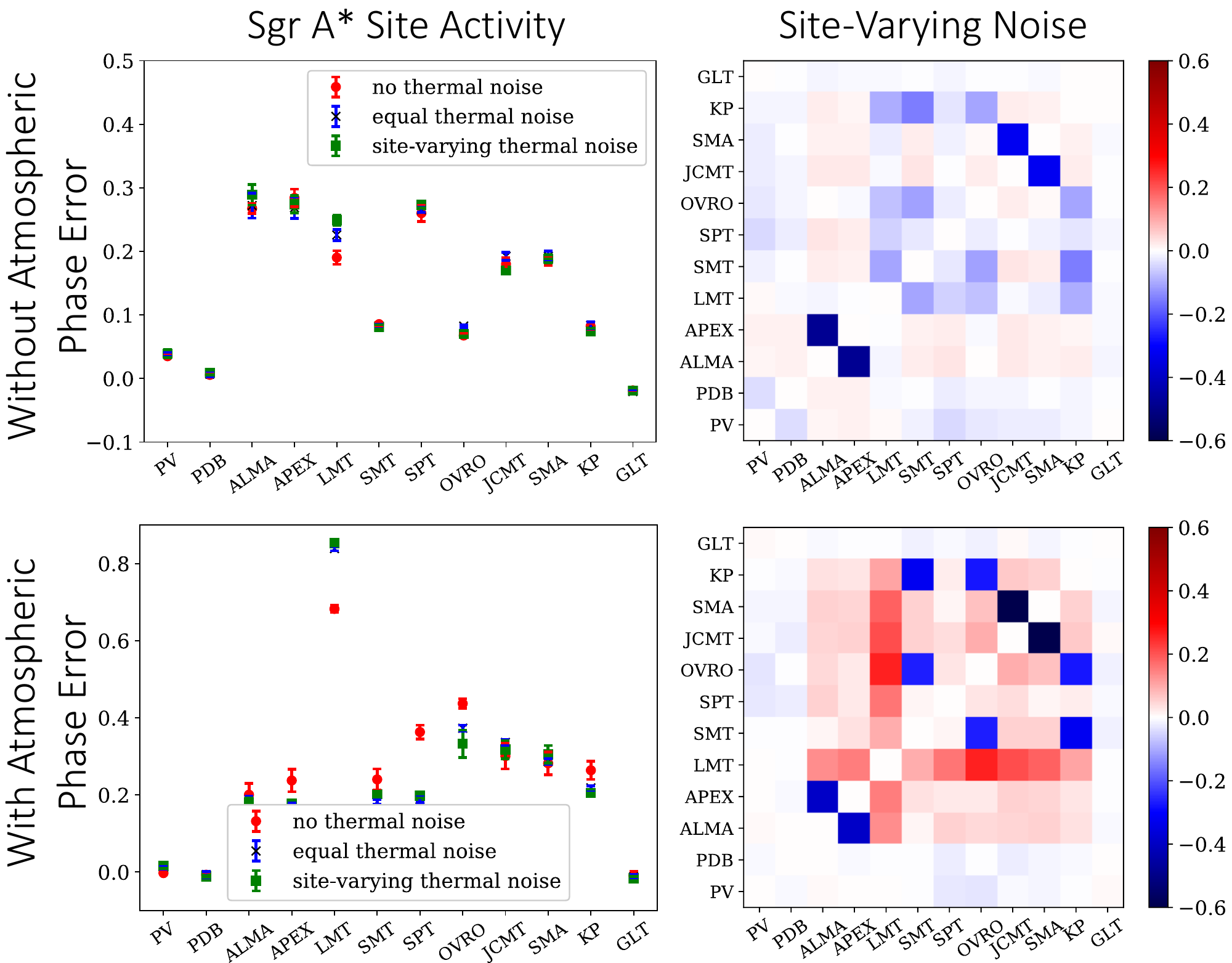}}
\caption{Effect of measurement noise on array design. Two rows respectively show the Ising model parameters without and with atmospheric phase error. The top row uses Reconstruction Network A from Fig.~\ref{fig:reconnet}, while the bottom row uses Reconstruction Network B to handle atmospheric noise. Thermal noise, no matter equal or site-varying, does not significantly change the learned telescope sensor-sampling. However, atmospheric phase noise is very impactful. According to a clique analysis in Sec.~\ref{subsubsec:noise}, nearly colinear telescopes are less favored after introducing atmospheric phase error. In both cases, only the correlation matrices for site-varying noise (cases (3) and (6) in Sec.~\ref{subsubsec:noise}) are shown.}
\label{fig:noise}
\end{figure}

Fig.~\ref{fig:noise} shows the learned Ising model parameters for these cases. As can be seen in the site activity plots, the sensor-sampling distributions are not significantly influenced by thermal noise only: (1), (2) and (3) are almost identical. However, the atmospheric noise significantly impacts telescope placement. With the presence of atmospheric phase error, LMT and OVRO become much more important, while SPT becomes less important. This effect is further amplified when both thermal noises and atmospheric errors appear. Short baselines, such as OVRO-LMT, are sampled more frequently, suggesting that it is especially important to include shorter baselines in high-resolution VLBI imaging when dealing with atmospheric phase errors.

Based on the undirected graph defined by the Ising model correlation matrix, we can also analyze all ``cliques" of size three to understand the importance of each closure triangle (i.e., closure phase measurement). In the affinity graph, telescopes are vertices and the pairwise correlations define edges. We define a subset of telescopes $(x_j, x_k, x_l)$ as a ``three-clique" if the correlations between each pair are significantly positive (larger than a threshold $\tau$), i.e.
\begin{equation}
\theta_{jk} > \tau, \theta_{kl} > \tau, \theta_{jl} > \tau.
\end{equation}
For instance, in case (6), eight three-cliques can be found in the corresponding graph given $\tau=0.04$. After ranking them with a simple metric,
\begin{equation}
    m_c =\theta_{jk}+\theta_{kl}+\theta_{jl},
\end{equation}
we find the five most important three-cliques are ``LMT-OVRO-JCMT", ``LMT-OVRO-SMA", ``APEX-LMT-JCMT", ``ALMA-LMT-JCMT" and ``LMT-JCMT-KP", all of which form triangles with similar baseline lengths (see Fig.~\ref{fig:map}). This suggests that closure phases from sites that are nearly colinear may not be as important for image reconstruction with atmospheric phase error.

\begin{figure*}[!t]
\centering
\setlength{\fboxrule}{0pt}
\framebox[\textwidth]{\includegraphics[width=0.75\textwidth]{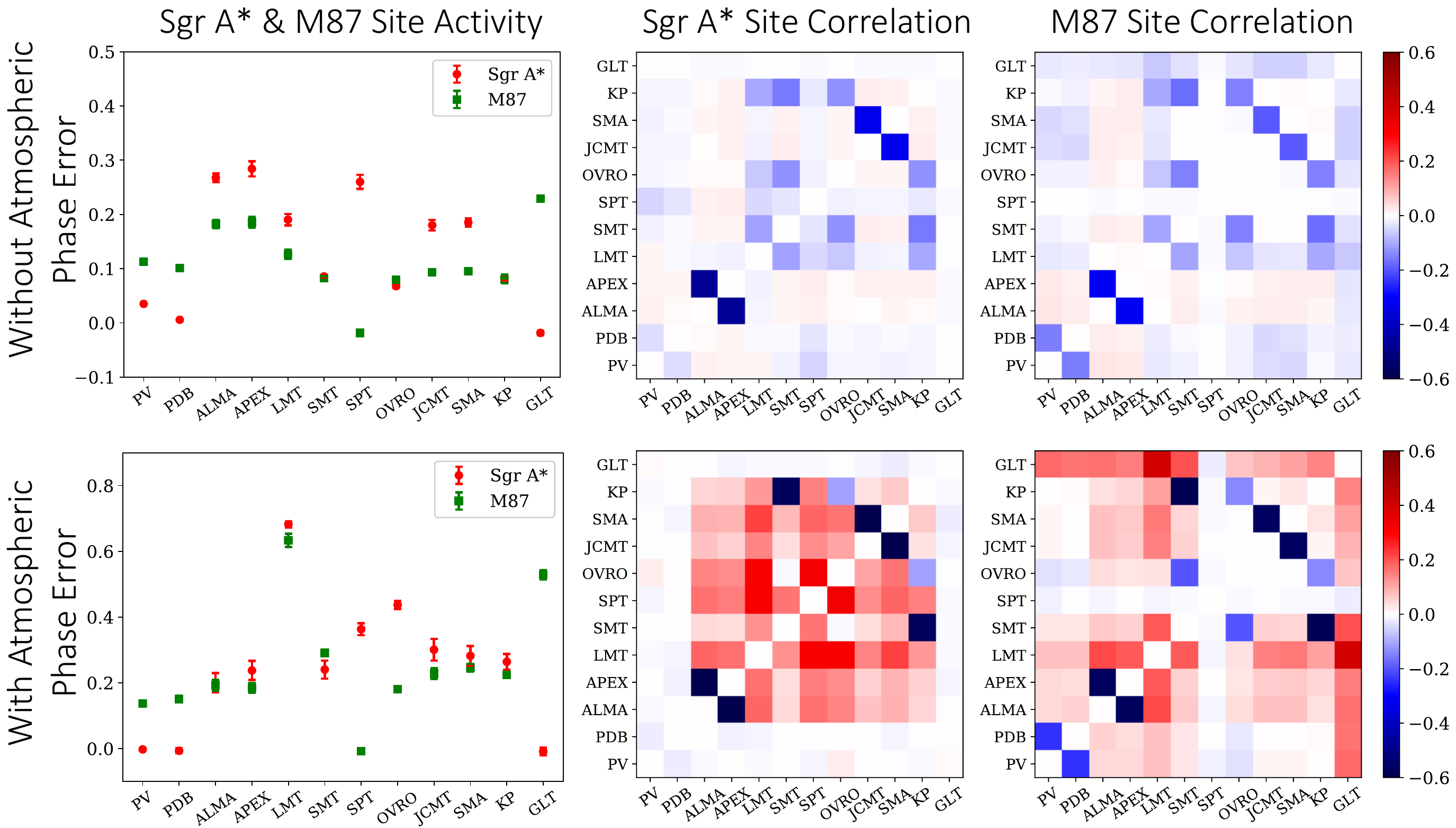}}
\caption{Comparison of the learned telescope importance for science targets Sgr A* and M87. Because the targets are at different declinations (Sgr A*: -29.24$^{\circ}$, M87: 12.39$^{\circ}$), different visibility measurements are sampled by the same telescope array (due to a change in the projected baselines). Therefore, different sets of telescopes are selected as important for the targets. For instance, GLT is very important for M87* but is the least important for Sgr A*, because it is occluded by Earth when observing targets in southern celestial hemisphere. The opposite is true for SPT. }
\label{fig:target}
\vspace{0.1in}
\centering
\setlength{\fboxrule}{0pt}
\framebox[\columnwidth]{\includegraphics[width=1.5\columnwidth]{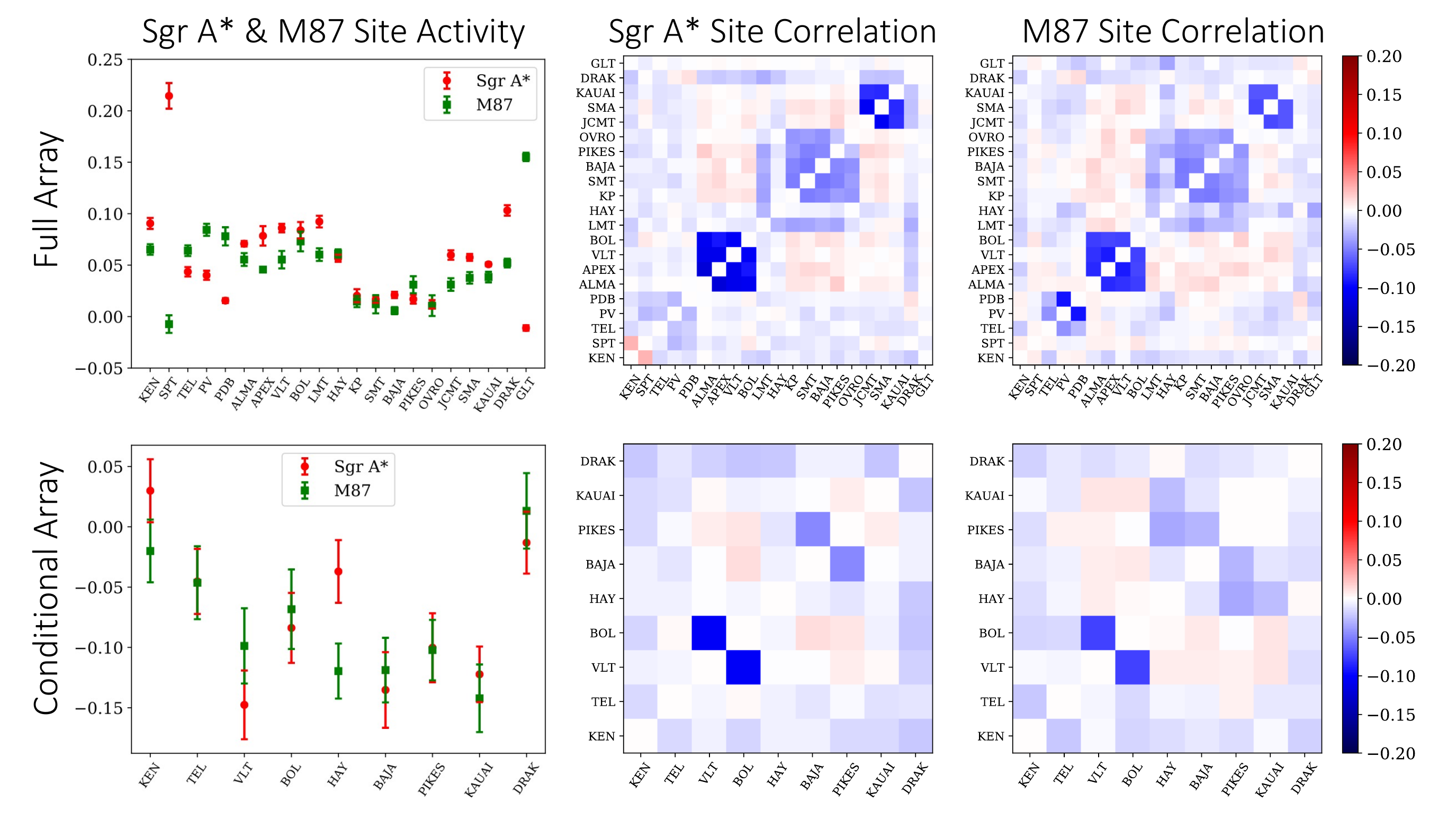}}
\caption{Ising model parameters for the ``FUTURE" array. The top figure shows the joint distribution of all sites in the ``FUTURE" array.
Negative correlations among close telescopes are strongly detected. The bottom row reports the Ising model distributions of nine additional sites conditioned on all the ``EHT+" telescopes being included in the array. 
We find that adding telescopes in Africa, particularly DRAK, would be particularly helpful in constraining both Sgr A* and M87* reconstructions. 
BAJA, which is close to KP, SMT and OVRO, has the least conditional activity. }
\label{fig:futurearray}
\end{figure*}

\subsubsection{EHT Design for Different Science Targets} \label{subsubsec:m87}
In this section, we compare the results of two different targets, Sgr A* and M87*. They are in the southern and northern celestial hemisphere, respectively (Sgr A* declination: -29.24$^{\circ}$, M87* declination: 12.39$^{\circ}$). Therefore Sgr A* and M87* should have different preferences for telescope site selections. Fig.~\ref{fig:target} reports the learned EHT sensor-sampling distributions considering no noise or only atmospheric phase noise. For both noise assumptions, SPT (occluded for M87* observations) becomes less important for M87* compared with Sgr A*, while PV, PDB and GLT become more important.

\subsubsection{Site Selection for ``FUTURE" array} \label{subsubsec:eht2025}
We study the site selection for a ``FUTURE" array, which includes nine additional telescope sites in addition to the twelve in ``EHT+". The same representative parameters described at the beginning of Sec.~\ref{subsec:experiment} are applied and no noise is introduced.
The learned Ising model parameters for both Sgr A* and M87 are shown in Fig.~\ref{fig:futurearray}. As can be seen, the results successfully capture the negative correlations among co-located telescopes, such as ALMA-APEX-VLT-BOL and JCMT-SMA-KAUAI.

Since the conditional distribution of a subset of Ising model elements is also an Ising model (shown in Eq.~\ref{eq:conditional}), we also report the conditional activities and correlations of nine additional sites assuming all the ``EHT+" telescopes are selected. Although nearly all nine sites have negative activities (due to the sparsity constraint of the original Ising model), the rank of these sites' importance still holds. We find that telescopes in Africa, particularly DRAK, would be helpful in constraining both Sgr A* and M87* reconstructions. However, we caution that more factors should be considered in the model before drawing conclusions about the importance of any one site, such as realistic atmospheric noise, weather assumptions, the science target's expected evolution, and both social and economic costs.

\subsubsection{Swapping Sampling Strategies} \label{subsubsec:switch}
In this section, we study the interaction of learned sampling strategies (encoders) and reconstruction methods (decoders) in order to demonstrate the advantage of co-design. In particular, we swap the sampling strategy learned for different types of noise (no noise $\&$ atmospheric phase noise) and science targets (Sgr A* $\&$ M87*), while keeping the reconstruction methods fixed. 
For instance, for two co-designed networks, we evaluate the quality of images reconstructed with the decoder of network 1, when the telescope measurements were sampled from the policy learned by network 2, and visa versa. 
To meaningfully compare networks, $\lambda_1$ and $\lambda_2$ were tuned so that the expected number of sampled telescopes in the EHT+ array for the different networks were nearly equal ($\approx$ 8).


Table~\ref{tab:swith} reports shift-invariant reconstruction losses (Eq.~\ref{eq:shiftinvariant}) of various ``recombined" autoencoders for a $res$ FWHM of 0.5 $\times$ the nominal beam. Each row represents a learned reconstruction method and each column represents a learned sensor sampling distribution. The reconstruction metric is computed based on 1000 test images in the ``MNIST" dataset. As expected, the optimal sampling strategy and reconstruction method are those that are learned jointly: a particular VLBI array sampling policy works best with its corresponding reconstruction method. 

\begin{figure}[h]
\centering
\setlength{\fboxrule}{0pt}
\framebox[\columnwidth]{\includegraphics[width=1.0\columnwidth]{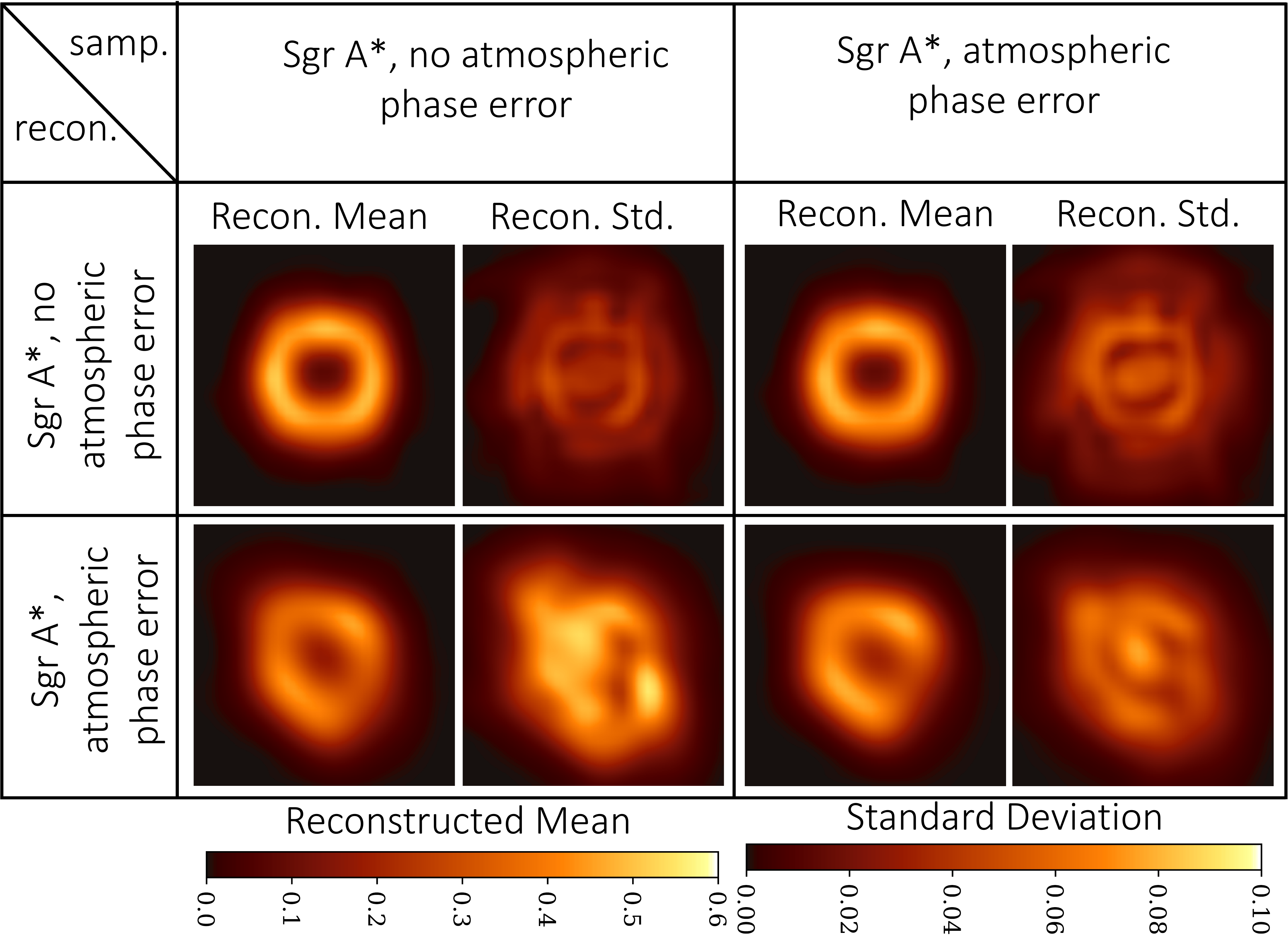}}
\caption{Reconstruction mean and standard deviation from 1000 sampling trials of the ``ring" image in Fig.~\ref{fig:blur_recon} using ``recombined" auto-encoders. Reconstruction performance is best for the co-designed networks; the reconstructed standard deviation is significantly higher for the ``recombined" autoencoders with independently learned encoders and decoders.    }
\label{fig:swap}
\end{figure}

Fig.~\ref{fig:swap} shows the mean reconstruction and standard deviation resulting from 1000 sampling trials of the same truth image (see Fig.~\ref{fig:blur_recon}) for each ``recombined" autoencoder. The standard deviation is significantly lower for autoencoders that were co-designed.
This simple tests help to confirm the co-designed sensor-sampling and reconstruction strategies outperform independently learned strategies.




\section{Discussion}
\label{sec:summary}

Optimal sensing is important for resource-limited image reconstruction. We presented an approach to learn an optimal sensor-sampling distribution for a computational imaging system.
This method optimizes an Ising sensor-sampling distribution jointly with an image reconstruction method, implemented in a physics-constrained,  fully  differentiable,  autoencoder.

We demonstrated the proposed framework on a VLBI telescope array design task, where sensor correlations and atmospheric noise present unique challenges.
Experiments show that the proposed technique can be helpful in planing future telescope array designs and developing observation strategies.
We also believe the proposed framework is applicable to a variety of other sensor selection/sampling problems in computational imaging, such as the optimization of LED array illumination patterns in Fourier ptychography~\cite{kellman2019data} and the sampling pattern in $\kappa$-space for fast-MRI~\cite{bahadir2019adaptive}.
Changing the application simply requires changing the physics-based forward model, $f(z)$, the decoder, $A_w\{\cdot\}$, in Eq.~\ref{eq:loupe} and the training data. By co-designing the Ising model sampling distribution simultaneously with the reconstruction decoder, we can better optimize the design of future computational imaging systems.

\begin{table}[!t]
\renewcommand{\arraystretch}{1.3}
\caption{Reconstruction losses after swapping sampling strategies. }
\centering
\begin{tabular}{|c|c|c|c|c|}
\hline
\diagbox[innerwidth=1.0cm]{recon.}{samp.} & \thead{Sgr A$^{\star}$, no\\atmospheric\\ phase error} & \thead{Sgr A$^{\star}$,\\ atmospheric\\ phase error} & \thead{M87, no\\atmospheric\\ phase error} & \thead{M87,\\ atmospheric\\ phase error} \\ 
\hline
\thead{Sgr A$^{\star}$, no\\atmospheric\\ phase error} & \textbf{0.0072} & 0.0088 & 0.0168 & 0.0165 \\
\hline
\thead{Sgr A$^{\star}$,\\ atmospheric\\ phase error} & 0.0199 & \textbf{0.0163} & 0.0252 & 0.0230 \\
\hline
\thead{M87, no\\atmospheric\\ phase error} & 0.0155 & 0.0164 & \textbf{0.0061} & 0.0079\\
\hline
\thead{M87,\\ atmospheric\\ phase error} & 0.0171 & 0.0152 & 0.0127 & \textbf{0.0116}\\
\hline
\end{tabular}
\label{tab:swith}
\end{table}

\hspace{7in}
\section*{Acknowledgments}
The authors would like to thank Lindy Blackburn, Alexander Raymond, Michael Johnson, and Sheperd Doeleman for helpful discussions on the constraints of a next-generation EHT array, and Michael Kellman for helpful discussions on Fourier ptychography.
This work was supported by NSF award 1935980: ``Next Generation Event Horizon Telescope Design," and Beyond Limits.

\bibliographystyle{IEEEtran}
\bibliography{references}

\ifpeerreview \else



\begin{IEEEbiography}[{\includegraphics[width=1in,height=1.25in,clip,keepaspectratio]{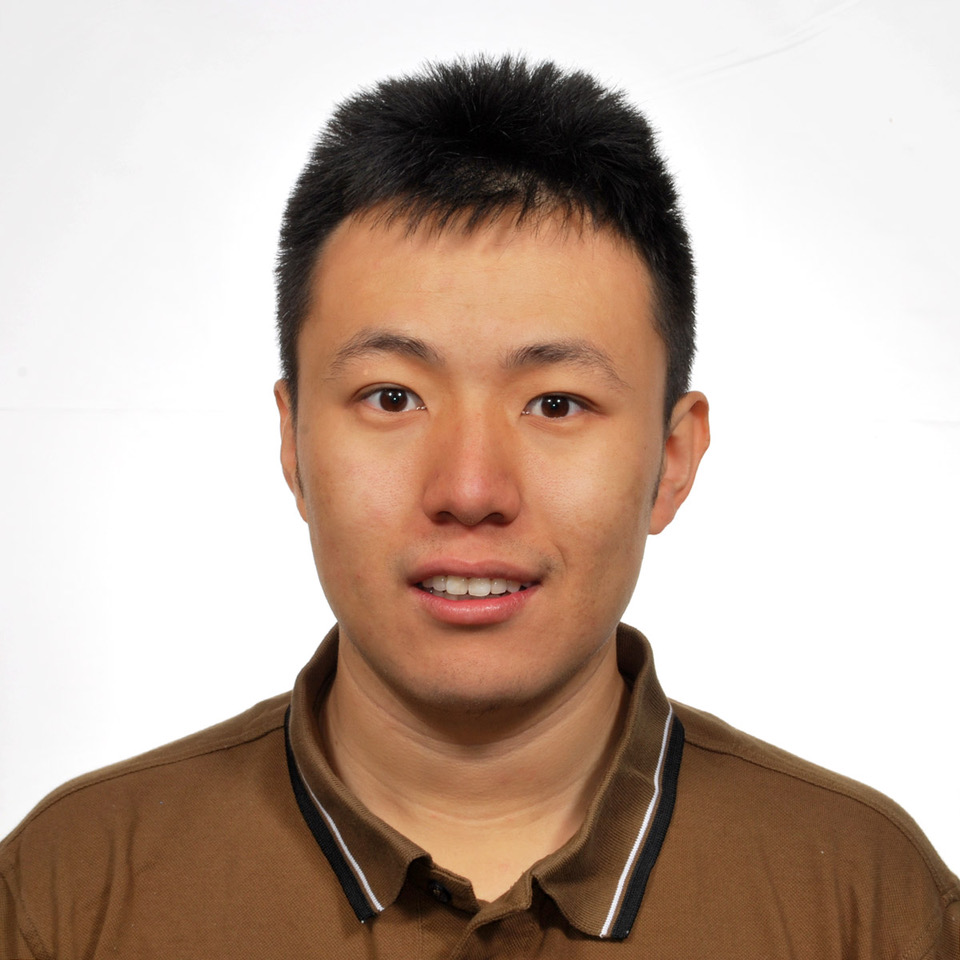}}]{He Sun}
is a postdoctoral researcher in Computing and Mathematical Sciences (CMS) at the California Institute of Technology. He received his PhD from Princeton University in 2019 and his BS degree from Peking University in 2014. His research focuses on adaptive optics and computational imaging, especially their applications in astrophysical and biomedical sciences.
\end{IEEEbiography}
\vskip -2\baselineskip plus -1fil
\begin{IEEEbiography}[{\includegraphics[width=1in,height=1.25in,clip,keepaspectratio]{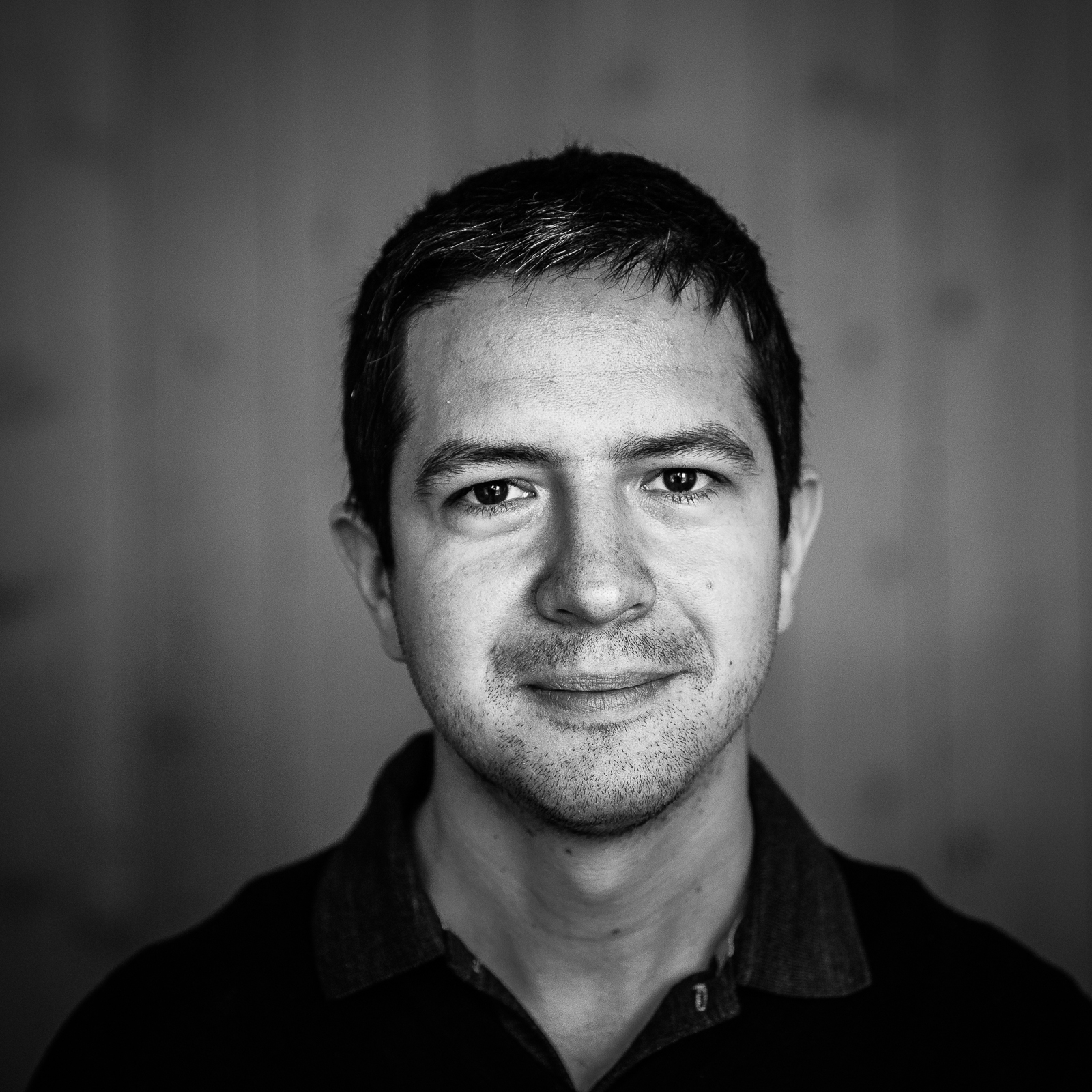}}]{Adrian V. Dalca}
is an assistant professor at Harvard Medical School, Massachusetts General Hospital, and a Research Scientist in CSAIL, Massachusetts Institute of Technology. His research focuses on machine learning techniques and probabilistic models for medical image analysis. Driven by clinical questions, he develops core learning algorithms, as well as registration, segmentation and imputation methods aimed at clinical-sourced datasets and broadly applicable in image analysis. \end{IEEEbiography}
\vskip -2\baselineskip plus -1fil
\begin{IEEEbiography}[{\includegraphics[width=1in,height=1.25in,clip,keepaspectratio]{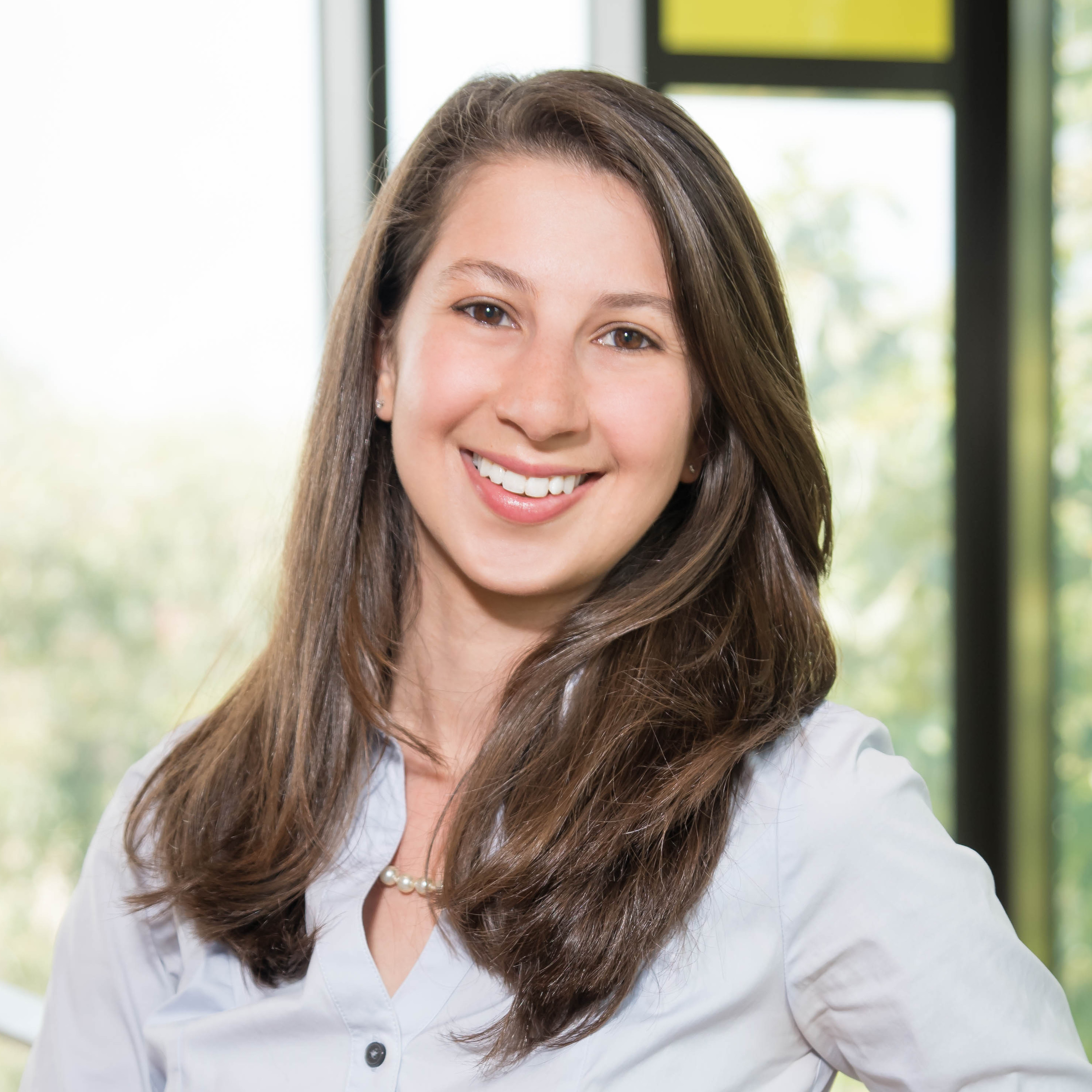}}]{Katherine L. Bouman}
is an assistant professor in Computing and Mathematical
Sciences (CMS) and Electrical Engineering (EE) and a Rosenberg Scholar at the California Institute of Technology (Caltech) in Pasadena,
California. 
Her group combines ideas from signal processing, computer vision, machine learning, and physics to design computational imaging systems. 
These systems tightly integrate algorithm and sensor design, making it possible to observe phenomena previously difficult or impossible to measure with traditional approaches. 
She was recognized as a co-recipient of the
Breakthrough Prize in fundamental physics, and was named the 2020 imaging “Scientist of the Year” by
the Society for Imaging Science and Technology for her work capturing the first picture of a black hole with the international Event Horizon Telescope Collaboration. 
\end{IEEEbiography}


\fi

\end{document}


\maketitle

\begin{table}[!t]
\renewcommand{\arraystretch}{1.3}
\caption{Additional telescope sites in "FUTURE" array. }
\centering
\begin{tabular}{c||c|c}
\hline
Sites & Location & SEFD\\
\hline\hline
HAY & Westford, Massachusetts & NA \\
\hline
PIKES & Pikes Peak, Colorado & NA \\
\hline
BAJA & National Astronomical Observatory, Mexico & NA\\
\hline
GAM & Gamsberg, Namibia & NA\\
\hline
KAUAI & Kaua'i, Hawaii & NA\\
\hline
KEN & Mt. Kenya, Kenya & NA\\
\hline
BOL & Chacaltaya Astrophysical Observatory, Bolivia & NA\\
\hline
VLT & Mt. Cerro Paranal, Chile & NA\\
\hline
DRAK & Drakensberg, South Africa& NA \\
\hline
\end{tabular}
\label{tab:futuresites}
\end{table}

\begin{figure}[!t]
\centering
\setlength{\fboxrule}{0pt}
\framebox[\columnwidth]{\includegraphics[width=1\columnwidth]{noise_effect_supplemental.pdf}}
\caption{Learning correlation matrices for cases (1) no noise, (2) equal thermal noise, (4) atmospheric phase noise (5) atmospheric phase noise + equal thermal noise in Sec.~4.3.4. Thermal noise causes little effect on the site correlations.}
\label{fig:noise_supp}
\end{figure}

